\documentclass[twocolumn]{aastex631}
\newcommand{\fo}{\ensuremath{f^\parallel}}
\newcommand{\fe}{\ensuremath{f^\perp}}

\newcommand{\pq}{\ensuremath{P_Q}}

\newcommand{\pu}{\ensuremath{P_U}}

\newcommand{\nnq}{\ensuremath{N_Q}}
\newcommand{\nnu}{\ensuremath{N_U}}

\newcommand{\ainv}{\ensuremath{\alpha_{0}} }
\newcommand{\amin}{\ensuremath{\alpha_{\rm{min}}} }
\newcommand{\amax}{\ensuremath{\alpha_{\rm{max}}} }

\newcommand{\pmin}{\ensuremath{P_{\rm{min}}} }
\newcommand{\pmax}{\ensuremath{P_{\rm{max}}} }

\usepackage{amsmath}
\usepackage{verbatim}
\usepackage{graphicx}
\usepackage{subfigure}
\usepackage{chngcntr} 
\usepackage{appendix}
\usepackage{textcomp}

\usepackage{chngcntr} 

\begin{document}

\title{Polarimetry of Didymos-Dimorphos: Unexpected Long-Term Effects of the DART Impact}

\author[0000-0002-6610-1897]{Zuri Gray}
\affiliation{Armagh Observatory \& Planetarium, College Hill, Armagh, BT61 9DG, UK \\}
\affiliation{Mullard Space Science Laboratory, Department of Space and Climate Physics, University College London, Holmbury St. Mary, Dorking, Surrey RH5 6NT, UK}
\affiliation{Nordic Optical Telescope, Rambla Jos\'{e} Ana Fernández P\'{e}rez 7, ES-38711 Bre\~{n}a Baja, Spain}
\thanks{Email: zuri.gray@armagh.ac.uk}

\author[0000-0002-7156-8029]{Stefano Bagnulo}
\affiliation{Armagh Observatory \& Planetarium, College Hill, Armagh, BT61 9DG, UK}
\affiliation{Mullard Space Science Laboratory, Department of Space and Climate Physics, University College London, Holmbury St. Mary, Dorking, Surrey RH5 6NT, UK}

\author[0000-0002-5624-1888]{Mikael Granvik}
\affiliation{Department of Physics, PO Box 64, FI-00014 University of Helsinki, Finland}
\affiliation{Asteroid Engineering Laboratory, Lule\r{a} University of Technology, Box 848, SE-98128 Kiruna, Sweden}

\author[0000-0002-6645-334X]{Alberto Cellino}
\affiliation{INAF, Osservatorio Astrofisico di Torino, I-10025 Pino Torinese, Italy}

\author[0000-0002-5859-1136]{Geraint H. Jones}
\affiliation{Mullard Space Science Laboratory, Department of Space and Climate Physics, University College London, Holmbury St. Mary, Dorking, Surrey RH5 6NT, UK}
\affiliation{The Centre for Planetary Sciences at UCL/Birkbeck, Gower Street, London WC1E 6BT, UK}

\author[0000-0002-9321-3202]{Ludmilla Kolokolova}
\affiliation{Department of Astronomy, University of Maryland, College Park, MD 20742-2421, USA}

\author[0000-0003-0670-356X]{Fernando Moreno}
\affiliation{Instituto de Astrof\'{\i}sica de Andaluc\'{\i}a, CSIC, Glorieta de la Astronomia s/n, E-18008 Granada, Spain}

\author[0000-0001-8058-2642]{Karri Muinonen}
\affiliation{Department of Physics, PO Box 64, FI-00014 University of Helsinki, Finland}

\author[0000-0002-5138-3932]{Olga Mu\~{n}oz}
\affiliation{Instituto de Astrof\'{\i}sica de Andaluc\'{\i}a, CSIC, Glorieta de la Astronomia s/n, E-18008 Granada, Spain}

\author[0000-0002-9298-7484]{Cyrielle Opitom}
\affiliation{Institute for Astronomy, University of Edinburgh, Royal Observatory, Edinburgh, EH9 3HJ, UK}

\author[0000-0001-7403-1721]{Antti Penttil\"a}
\affiliation{Department of Physics, PO Box 64, FI-00014 University of Helsinki, Finland}

\author[0000-0001-9328-2905]{Colin Snodgrass}
\affiliation{Institute for Astronomy, University of Edinburgh, Royal Observatory, Edinburgh, EH9 3HJ, UK}


\received{29 August 2023}
\accepted{21 November 2023}

\begin{abstract}
We have monitored the Didymos-Dimorphos binary system in imaging polarimetric mode before and after the impact from the Double Asteroid Redirection Test (DART) mission. A previous spectropolarimetric study showed that the impact  caused a dramatic drop in polarisation. Our longer-term monitoring shows that the polarisation of the post-impact system remains lower than the pre-impact system even months after the impact, suggesting that some fresh ejecta material remains in the system at the time of our observations, either in orbit or settled on the surface. The slope of the post-impact polarimetric curve is shallower than that of the pre-impact system, implying an increase in albedo of the system. This suggests that the ejected material is composed of smaller and possibly brighter particles than those present on the pre-impact surface of the asteroid. Our polarimetric maps show that the dust cloud ejected immediately after the impact polarises light in a spatially uniform manner (and at a lower level than pre-impact). Later maps exhibit a gradient in polarisation between the photocentre (which probes the asteroid surface) and the surrounding cloud and tail. The polarisation occasionally shows some small-scale variations, the source of which is not yet clear. The polarimetric phase curve of Didymos-Dimorphos resembles that of the S-type asteroid class. 
\end{abstract}


\section{Introduction} 
The Double Asteroid Redirection Test (DART) mission aimed to study and demonstrate the effectiveness of the kinetic impact technique for deflecting potentially hazardous asteroids and, in doing so, estimate the momentum transfer efficiency ($\beta$) of the impact \citep{daly2023}. At 23:14 UT on September 26th 2022, the DART spacecraft successfully collided with its target, Dimorphos -- the moonlet of near-Earth asteroid, Didymos. Following the impact, observations revealed a greater-than-expected change in Dimorphos' orbital period \citep{thomas2023}, as well as the formation of a dust cloud and tail due to material ejection \citep{li2023,opitom2023}. The DART mission provided a rare opportunity to study the characteristics and behaviour of an ejecta cloud, which is essential for enhancing our understanding of the global properties of asteroids and the dynamics of their collisions, thus, improving our ability to assess and mitigate potential future asteroid threats.

A worldwide observing campaign was initiated to observe this once in a lifetime event using both ground-based and space telescopes, with polarimetric observations being a crucial aspect. Polarimetry exploits the fact that sunlight scattered by the surface of astronomical bodies becomes partially polarised. For Solar System objects, the degree of linear polarisation is measured as a function of phase angle $\alpha$, the angle between the Sun and observer as seen from the target. It is found that all small Solar System objects show a similar polarimetric phase angle dependence: the polarisation is zero at zero phase angle, negative at small phase angles, zero at the so-called inversion angle $\ainv$, and then positive at larger phase angles. Thus, the polarimetric phase curve consists of the negative polarisation branch (NPB) and the positive polarisation branch (PPB). The concept of positive and negative polarisation refers to the reference direction that we adopt to measure it: the flux perpendicular to the scattering plane (the Sun-object-observer plane) minus the flux parallel to that plane, divided by the sum of the two fluxes. The polarisation is said to be positive when it is mostly oriented in the direction perpendicular to the scattering plane, and vice versa for negative polarisation. 

Despite general similarities, the overall shape of the polarimetric phase curve will depend on the global properties unique to each object, including the material composition, size distribution, shape and roughness \citep{shkuratov2007,escobar2017,munoz2021}. Thus, certain parameters describing this curve, including the minimum and maximum values of polarisation, \pmin and $\pmax$, and the phase angles at which they occur, \amin and $\amax$, as well as the inversion angle and the slope of the curve at this angle \textit{h}, will vary from object to object. It is worth noting that \pmax\ and \amax\ are rarely measured for asteroids. Due to the orbital path and observing geometry of main-belt asteroids, the observable phase angle range is typically $\sim 0-30^{\circ}$. Only some near-Earth objects (NEOs) can be explored at larger phase angles when they pass sufficiently close to Earth.  

In this study, we report our extensive imaging polarimetric observations of the Didymos-Dimorphos system. These observations allow us to determine the polarimetric behaviour of the system in its original state (pre-impact) and observe how it changes following the impact of the DART spacecraft (post-impact). We also study and monitor the spatial and temporal evolution of the ejecta dust cloud and tail in polarised light. An advantage of the use of polarimetry compared to other techniques which rely heavily of intensity alone is the fact that polarisation is measured as a ratio, i.e., a percentage. This means it is not necessary to normalise the data in brightness in order to compare data taken of the same object at different times. In this paper, we present both qualitative and quantitative results. Further interpretation of these results, as well as possible inter-data comparisons, will follow in subsequent papers. 


\section{Observations}\label{Sect_Observations}
Imaging polarimetric observations of the Didymos-Dimorphos system were obtained with FORS2 (FOcal Reducer/low dispersion Spectrograph 2) at the VLT (Very Large Telescope) and ALFOSC (Alhambra Faint Object Spectrograph and Camera) at the NOT (Nordic Optical telescope), with specific details provided in Table \ref{tab:obslog}. FORS2 is a multipurpose instrument capable of performing low-resolution spectroscopic and imaging observations, as well as polarimetry when equipped with the appropriate optics. Following the design first suggested by \cite{appenzeller1967}, the polarimetric setup of FORS2 consists of a rotatable retarder waveplate followed by a Wollaston prism. The retarder waveplate introduces a phase-shift between the orthogonal polarisation components of the light (ordinary and extraordinary) while the Wollaston prism splits the light into these two components. In imaging polarimetric mode, a Wollaston mask is used to prevent the superposition of the beams split by the prism. The field of view (FoV) is vignetted by the mask consisting of nine $6.8' \times 22''$ strips. Approximately half of the observations were performed with 
$1\!\times\!1$ binning and the other half with $2\!\times\!2$ binning, with a pixel scale of $0.125 " \rm{pixel}^{-1}$ and $0.25 " \rm{pixel}^{-1}$, respectively. This instrument setup allows the user to measure the degree of polarisation using the so-called beam swapping technique, which largely suppresses the effects of instrumental polarisation \citep[e.g.][]{bagnulo2009}. For this, observations were taken with the retarder-waveplate ($\lambda/2$ waveplate in this case) at eight different position angles between $0^{\circ} - 157.5^{\circ}$ increasing in steps of $22.5^{\circ}$. In some cases, however, observations were taken with the retarder-waveplate in position angles separated by increments of $45^{\circ}$ only, and with the instrument position angle set equal to the angle perpendicular to the scattering plane (assuming that in the instrument reference system, Stokes $U$ would be zero for symmetry reasons).

Linear polarimetry with ALFOSC is performed with a similar set up to that of FORS2: a retarder waveplate is mounted in the FAPOL unit and a calcite plate is mounted in the aperture wheel. The main difference, however, is the lack of a Wollaston mask in ALFOSC, which prevents us from measuring the polarisation of extended objects. The calcite plates produces a vignetted field of about $2.7'$ in diameter and the observed object is separated by about $15"$. During each NOT observing night, two to four sets of observations were performed in each filter with $2\!\times\!2$ binning, with the ($\lambda/2$) retarder-waveplate rotated at sixteen different position angles between $0^{\circ} - 337.5^{\circ}$ in steps of $22.5^{\circ}$. The final value of polarisation of a given night is the weighted average of the set of observations of that night. Finally, we observed high polarisation stars, BD+64106 and HD204827, and zero polarisation star, BD+284211, on a number of the NOT observing nights. Comparing our measurements to literature values, we calculated the instrumental polarisation in both the R- and B-filters, which we used to calibrate our data. This step was not necessary for VLT observations as measurements of standard stars are already part of the calibration plan of the instrument. 

The DART spacecraft impacted Dimorphos at 23:14 UT on September 26th, 2022. Our observations span a period from about one month before to almost four months after the impact, with a number of the VLT observations being made hand-in-hand with spectropolarimetric observations \citep{bagnulo2023}---this has been indicated in the observing log. Considering the entire observing campaign, we have observed the asteroid system in three stages. (i) Pre-impact, when the system remained unperturbed and the surface properties are those resulting from the previous collisional history of the system. (ii) Immediate post-impact, characterised by a large ejected dust cloud that persisted for several weeks, influencing polarimetric measurements. As time passed, the cloud dispersed, giving rise to an extensive tail, marking the third and final stage of observation. (iii) Long-term post-impact, possibly representing a ``new" system to that of its pre-impact state. 

Throughout the campaign, we observed Didymos-Dimorphos on six pre-impact and 29 post-impact nights using various filters and covering a phase angle range of approximately $7-76^{\circ}$. With FORS2, we used the broadband filters b\_HIGH, v\_HIGH, R\_SPECIAL, and I\_BESS (B-, V-, R-, and I-filter hereafter) centred around 440\,nm, 557\,nm, 655\,nm and 768\,nm, respectively. With ALFOSC, we used the B\_BES and R\_BES filters (B- and R- hereafter) centred around 440\,nm and 650\,nm. In total, we have accumulated 148 measurements: 67 in the B filter, 4 in the V filter, 74 in the R filter, and 5 in the I filter. This large number is a consequence of the observations being repeated 2-4 times in each filter during each night with NOT. As mentioned, the final recorded value of polarisation is calculated as the weighted average of these repeated observations. Therefore, we have ended up with 30 measurements in the B filter and 33 in R. 
\section{Data Reduction} \label{Sect_Reduction}

\subsection{Data Pre-Processing}
All data reduction processes were performed using dedicated, original Python scripts. Prior to any scientific reduction, bias and flat-field corrections were applied to all frames. For each observing night, a master-bias image was generated from a series of zero-exposure frames and a master-flat was created for each filter using the median of a series of exposures of a uniform light source. Twilight (sky) flats were used for FORS2 while dome flats were used in the case of ALFOSC. For each observing night, the corresponding master-bias was subtracted from each science frame, which was then divided by its corresponding master-flat image. To ensure data quality, all files were carefully inspected and those in which the asteroid photocentre passed in front of a background star were discarded, as well as their corresponding pair.

\begin{figure*}
\begin{center}
    \centering
    \includegraphics[width=\textwidth]{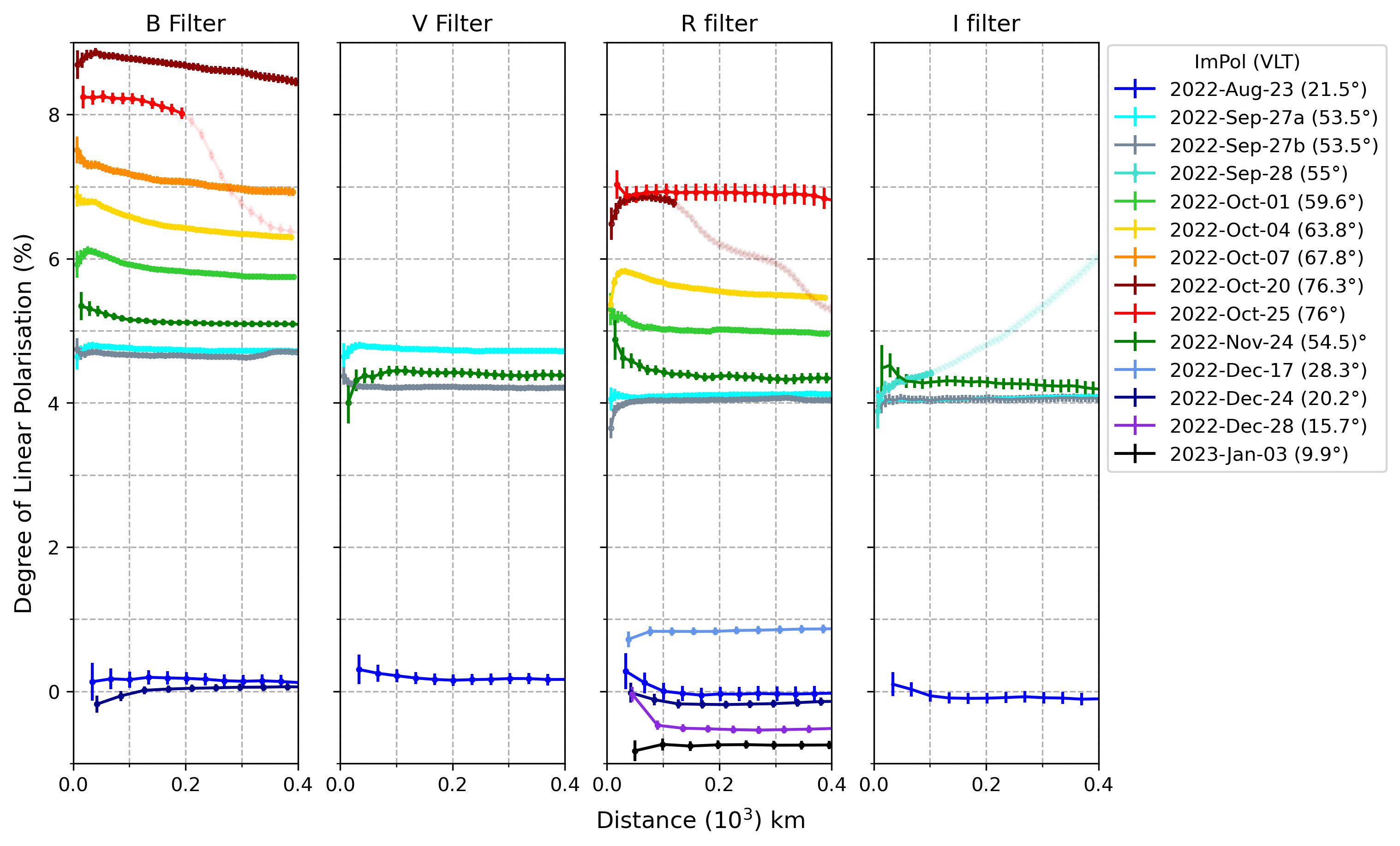}
    \caption{The polarimetric growth curves of imaging polarimetric VLT data of Didymos-Dimorphos in BVRI filters, i.e, the polarisation is measured in apertures of increasing size (pixel radii, where 0.125``/pixel). The colour of each curve represents an observing epoch and phase angle, indicated on the right. The pixel scale varied from epoch to epoch due to the changing asteroid-observer distance ($\Delta$). Thus, all curves have been scaled to the same distance scale (horizontal axis). The values reported in Table \ref{tab:obslog} and Figure \ref{fig:phasecurve} are those integrated in an aperture radius equivalent to 100 km (0.1 ($10^3$ km) on the x-axis). A number of the curves are spurious at larger radii due to the contamination of background stars: 2022-Oct-25 in the B filter, 2022-Oct-20 in the R filter, and 2022-Sep-28 in the I filter. These spurious signals have been left unshaded.}
    \label{fig:growthcurves}
\end{center}
\end{figure*}

\subsection{Polarimetry} 
To study the polarimetric behaviour of Didymos-Dimorphos, we measured the reduced Stokes parameters \citep{shurcliff1962}, \pq\ and \pu\, using the so-called beam-swapping technique \citep{bagnulo2009}. For this, we first measure the reduced Stokes parameters in the instrument reference system as:
\begin{equation}
\label{eq:pqpu}
\begin{split}
    \pq' &= \frac{1}{2} [D (\phi = 0^{\circ}) + D (\phi = 90^{\circ})], \\
    \pu' &= \frac{1}{2} [D (\phi = 22.5^{\circ}) + D (\phi = 112.5^{\circ})],
\end{split}
\end{equation}
where 
\begin{equation}
D (\phi) = \left( \frac{f^{\parallel} - f^{\perp}}{f^{\parallel} + f^{\perp}} \right) _{\phi} - \left( \frac{f^{\parallel} - f^{\perp}}{f^{\parallel} + f^{\perp}} \right)_{\phi+45\degr}
\end{equation}
and \fe\ and \fo\ are the fluxes measured in the perpendicular and parallel beam, respectively, obtained with the retarder waveplate at the position angle $\phi$. Where applicable, we transformed the Stokes parameters into the reference direction perpendicular to the scattering plane, and hence, the reference system of the target using the equation: 

\begin{equation}
\begin{split}
\pq &=\phantom{-} \cos(2\chi)\pq' + \sin(2\chi)\pu',  \\
\pu &= -\sin(2\chi)\pq' + \cos(2\chi)\pu', 
\end{split}
\label{eq:trans}
\end{equation}
\noindent
Here, $\chi = {\rm PA} + \Phi + 90^{\circ}$ and $\Phi$ is the angle between the asteroid-north celestial pole direction and the asteroid-Sun direction. After this transformation, we expect all the polarisation to be in \pq\ while \pu\ is expected to be zero within uncertainties for symmetry reasons. Further, the Null \nnq\ and \nnu\ parameters, which are calculated by subtracting the second $D(\phi)$ expression in Eq. \ref{eq:pqpu} rather than adding it, can be used as as an additional quality check for our measurements. Again, the Null parameters should be zero within uncertainty \citep{bagnulo2009}.

In this study, we performed aperture polarimetry and generated polarimetric maps. In both cases, the initial step was to accurately determine the position of the asteroid photocentre in both the \fe\ and \fo\ beams. This was accomplished using the DAOStarFinder Python package, which uses point spread function (PSF) estimation for source detection. Next, we measured the flux of the target in the parallel and perpendicular beams in a circular aperture, while the background sky level was measured in a region of the CCD free from background stars and the dust cloud and tail resulting from the impact. The \fo\ and \fe\ fluxes of the frames of each position of the retarder waveplate ($0, 22.5^{\circ}$...), once corrected for background sky, were then combined using Eq. \ref{eq:pqpu} and transformed with Eq. \ref{eq:trans} to obtain the final values of the reduced Stokes parameters. 

\subsection{Maps}
With the VLT data, we generated both polarimetric and imaging maps. The polarimetric maps were obtained in a similar manner to that of aperture polarimetry, but this time, the entire \fe\ and \fo\ beams, background subtracted, were combined (while aperture polarimetry extracts a numerical value only). The photocentres of the two beams of each waveplate angle ($0, 22.5, 45^{\circ}$, etc.) were aligned and merged using Eq. \ref{eq:pqpu} and \ref{eq:trans}. The resulting maps allow one to view the spatial distribution of the polarisation of the asteroid system. In the context of the DART mission, these maps are of particular interest as they make it possible to investigate small scale structures within the ejected dust cloud and tail, and track their spatial and temporal evolution. It is important to note that even minor misalignments, in the order of a few pixels, can give rise to spurious structures near the photocentre which may be misinterpreted as real variations in polarisation. Similarly, we co-added the two beams of all waveplate angles to create high S/N (signal-to-noise) imaging maps. These deep images provide a detailed view of the asteroid system and, hand-in-hand with the polarimetric maps, contribute to a better understanding of the overall features and characteristics of the ejected material.

\section{Results}\label{Sect_Results}

\subsection{Aperture Polarimetry}

\begin{table*}

\caption{\label{tab:obslog} Observing log. $\Delta$T indicates the time to/from the DART impact, $\Delta$ is the asteroid-observer distance, the scale is the km covered by 1" at each $\Delta$ value, $\alpha$ is the phase angle of the asteroid system at the time of the observation, INSTR is the instrument used for each observation, and \pq\ is the Stokes $Q$ parameter measured in the the BVRI filters, rotated to the direction perpendicular to the scattering plane. *VLT indicates spectropolarimetric measurements published in \cite{bagnulo2023} which we have included for completeness.}
 \begin{small}
\begin{center}
\begin{tabular}{cccccrr@{$\pm$}lr@{$\pm$}lr@{$\pm$}lr@{$\pm$}l}
\hline\hline    
DATE & $\Delta$T & $\Delta$ & Scale & $\alpha$ & INSTR & \multicolumn{8}{c}{$P_Q$(\%)} \\
YYYY-MM-M-DD & & (AU) & (km) & $(^{\circ})$ & & \multicolumn{2}{c}{B} & \multicolumn{2}{c}{V} & \multicolumn{2}{c}{R} & \multicolumn{2}{c}{I} \\ 
\hline   
2022-Aug-19 & -38d 20h 39m &  0.20 &  145  &  20.68 & NOT & \multicolumn{4}{c}{}               &-0.19  &  0.08 & \multicolumn{2}{c}{} \\
2022-Aug-23 & -34d 21h 28m &  0.18  &  130  &  21.50 & VLT   & 0.18 & 0.08 & 0.17 & 0.07          & -0.05 & 0.09 & -0.1 & 0.06 \\
2022-Aug-28 & -29d 20h 25m &  0.16  &  116  & 23.29 & *VLT   & 0.34 & 0.05 & 0.27 & 0.05          & 0.24 & 0.05 & 0.24 & 0.05 \\
2022-Sep-07 & -19d 15h 44m & 0.12 &  87 & 29.86 & *VLT   & 1.26 & 0.05 & 1.15 & 0.05          & 1.07 & 0.05 & 1.04 & 0.05 \\
2022-Sep-09 & -17d 21h 02m &  0.12 &  87 & 31.41 & NOT & 1.42  &  0.36  & \multicolumn{2}{c}{} &  1.25  &  0.09 & \multicolumn{2}{c}{} \\
2022-Sep-13 & -13d 20h 44m &  0.11 &  79 & 35.34 & NOT & 2.27  &  0.25  & \multicolumn{2}{c}{} &  1.7  &  0.08 & \multicolumn{2}{c}{} \\
2022-Sep-17 & -9d 16h 54m & 0.09 &  65  & 39.99 & *VLT   & 2.89 & 0.05 & 2.59 & 0.05          & 2.47 & 0.05 & 2.46 & 0.05 \\
2022-Sep-21 & -05d 20h 04m &  0.09 &  65 & 44.91 & NOT &4.03  &  0.1  & \multicolumn{2}{c}{} &  3.16  &  0.05 & \multicolumn{2}{c}{}  \\
2022-Sep-23 & -03d 18h 56m &  0.08 &  58 & 47.68 & NOT & 4.27  &  0.09  & \multicolumn{2}{c}{} &  3.55  &  0.05 & \multicolumn{2}{c}{}  \\
		     & -03d 19h 49m & & & & *VLT   & 4.30 & 0.05 & 3.79 & 0.05          & 3.61 & 0.05 & 3.59 & 0.05 \\
2022-Sep-26 &     -15h 49m & && 52.25 & *VLT   & 5.24 & 0.05 & 4.63 & 0.05          & 4.39 & 0.05 & 4.35 & 0.05 \\

\hline 
2022-Sep-27 &     +04h 47m &  0.08 &  58 & 53.48 & VLT  & 4.76 & 0.03 & 4.30 & 0.03          & 4.09 & 0.02 & 4.04 & 0.02 \\
		     &     +05h 37m &          &       &    & *VLT   & 4.56 & 0.05 & 4.16 & 0.05          & 3.98 & 0.05 & 4.01 & 0.05 \\
           	     &     +09h 54m &          &         &  & VLT   & 4.65 & 0.03 & 4.22 & 0.02          & 4.04 & 0.04 & 4.03 & 0.04 \\
2022-Sep-28 & +01d 09h 59m & 0.08 &  58 &  55.04 & *VLT   & 4.88 & 0.05 & 4.43 & 0.05          & 4.25 & 0.05 & 4.27 & 0.05 \\
		     & +01d 10h 30m &        &        &     & VLT  & \multicolumn{6}{c}{}                             & 4.41 & 0.04 \\        
2022-Sep-30 & +03d 04h 38m &  0.07 &  50 & 58.21 & NOT & 5.34  &  0.14  & \multicolumn{2}{c}{} &  4.59  &  0.06 & \multicolumn{2}{c}{}\\
		     & +03d 08h 30m &        &         &    & *VLT   & 5.39 & 0.05 & 4.86 & 0.05          & 4.65 & 0.05 & 4.65 & 0.05 \\
2022-Oct-01 & +04d 05h 36m &  0.07 &  50 & 59.53 & NOT &  5.63  &  0.11  & \multicolumn{2}{c}{} &  4.79  &  0.05 & \multicolumn{2}{c}{}\\
       		    & +04d 08h 59m &          &      &      & *VLT   & 5.63 & 0.05 & 5.1 & 0.05           & 4.87 & 0.05 & 4.84 & 0.05 \\
           	    & +04d 09h 19m &          &       &     & VLT  & 5.9 & 0.03 & \multicolumn{2}{c}{}  & 5.02 & 0.03 & \multicolumn{2}{c}{}\\
2022-Oct-02 & +05d 05h 12m &  0.07 &  50 & 61    & NOT & 5.96  &  0.07  & \multicolumn{2}{c}{} &  4.96  &  0.03 &\multicolumn{2}{c}{}\\
2022-Oct-03 & +06d 06h 17m &  0.07 &  50 & 62.43  & NOT &  6.24  &  0.18  & \multicolumn{2}{c}{} &  5.18  &  0.04 & \multicolumn{2}{c}{}\\
2022-Oct-04 & +07d 05h 14m &  0.07 &  50 & 63.83 & NOT & 6.65  &  0.11  & \multicolumn{2}{c}{} &  5.35  &  0.06 & \multicolumn{2}{c}{}\\
           & +07d 08h 21m &       &   &    & VLT   & 6.60 & 0.03 & \multicolumn{2}{c}{} & 5.67 & 0.03 & \multicolumn{2}{c}{}\\
2022-Oct-07 & +10d 09h 17m &  0.07 &  50 & 67.75 & VLT   & 7.17 & 0.04 & \multicolumn{6}{c}{}     \\
2022-Oct-08 & +11d 04h 58m &  0.07 &  50 & 68.86 & NOT &  7.25  &  0.12  & \multicolumn{2}{c}{} &  6.21  &  0.05 & \multicolumn{2}{c}{}\\
2022-Oct-11 & +14d 05h 28m &  0.07 &  50 & 71.87  & NOT & 8.18  &  0.19  & \multicolumn{2}{c}{} &  6.6  &  0.06 & \multicolumn{2}{c}{}\\
2016-Oct-17 & +20d 04h 38m &  0.08 &  58 & 75.56  & NOT & 8.38  &  0.29  & \multicolumn{2}{c}{} &  6.84  &  0.14 & \multicolumn{2}{c}{}\\
2022-Oct-20 & +23d 06h 54m &  0.09 &  65 & 76.27 & VLT   & 8.75 & 0.04 & \multicolumn{2}{c}{}  & 6.82 & 0.05 & \multicolumn{2}{c}{}\\
\vspace{0.3cm}
2022-Oct-21 & +24d 04h 49m &  0.09 &  65  & 76.34 & NOT & 8.83  &  0.83  & \multicolumn{2}{c}{} &  6.99  &  0.18 & \multicolumn{2}{c}{}\\
2022-Oct-24 & +26d 09h 01m &  0.09 &  65  & 76.16 & *VLT   & 9.08 & 0.05 & 7.91 & 0.05           & 7.47 & 0.05 & 7.37 & 0.05 \\
2022-Oct-25 & +28d 08h 44m &  0.10 &  72 & 75.98 & VLT   & 8.2 & 0.06 & \multicolumn{2}{c}{}   & 6.93 & 0.1 &  \multicolumn{2}{c}{}\\
2022-Oct-26 & +29d 06h 55m &  0.10 &  72 & 75.73 & NOT & 8.87  &  0.33  & \multicolumn{2}{c}{} &  7.13  &  0.3 & \multicolumn{2}{c}{}\\
2022-Oct-31 & +34d 05h 02m &  0.11 &  79 & 73.85 & NOT & 8.77  &  0.2  & \multicolumn{2}{c}{} &  7.04  &  0.11 & \multicolumn{2}{c}{}\\
2022-Nov-16 & +50d 03h 46m &  0.14 &  101 & 62.36 & NOT & 6.54  &  0.28  & \multicolumn{2}{c}{} &  5.03  &  0.12 & \multicolumn{2}{c}{}\\
2022-Nov-24 & +58d 07h 26m &  0.16 &  116 & 54.52 & VLT   & 5.14 & 0.04 & 4.46 & 0.06 & 4.42 & 0.05 & 4.28 & 0.06 \\
2022-Dec-02 & +66d 02h 47m &  0.18 & 130 & 46.09 & NOT & 3.68  &  0.22  & \multicolumn{2}{c}{} &  3.06  &  0.07 & \multicolumn{2}{c}{} \\
2022-Dec-16 & +80d 03h 12m &  0.21 & 152 & 29.73 & NOT & 1.24  &  0.28  & \multicolumn{2}{c}{} &  0.93  &  0.09 & \multicolumn{2}{c}{}\\
2022-Dec-17 & +81d 07h 13m &  0.21 & 152 & 28.29 & VLT   & \multicolumn{4}{c}{}               & 0.83 & 0.05 & \multicolumn{2}{c}{} \\
2022-Dec-24 & +88d 08h 55m &  0.23 & 166 & 20.17 & VLT   & 0.01 & 0.05 & \multicolumn{2}{c}{} & -0.18 & 0.05 & \multicolumn{2}{c}{} \\
2022-Dec-28 & +92d 06h 55m &  0.25 & 181 & 15.72 & VLT   & \multicolumn{4}{c}{}               & -0.51 & 0.05 & \multicolumn{2}{c}{} \\
2023-Jan-03 & +98d 08h 09m &  0.27 & 195 & 9.88 & VLT   & \multicolumn{4}{c}{}               & -0.76 & 0.06 & \multicolumn{2}{c}{} \\
2023-Jan-11 & +106d 05h 01m & 0.31 & 224 & 6.55 & NOT & -0.52  &  0.6  & \multicolumn{2}{c}{} &  -0.47  &  0.24 & \multicolumn{2}{c}{} \\
2023-Jan-19 & +114d 03h 09m & 0.36 & 261 & 10.41  & NOT & -0.82  &  0.37  & \multicolumn{2}{c}{} &  -0.83  &  0.18 & \multicolumn{2}{c}{}\\
\hline \hline
\end{tabular}
\end{center}
 \end{small}
\end{table*}

Figure \ref{fig:growthcurves} shows the polarisation measured as a function of aperture radius increasing in steps of one pixel, i.e., the polarimetric growth curves, of the VLT data. Each data point of these curves represents an increase of one pixel in aperture radius. The asteroid-observer distance ($\Delta$), and hence, the pixel scale varied throughout the observing campaign. Taking this into account, we have scaled all growth curves to the same distance scale. The plot shows a mix of both flat and slightly varying growth curves. In the case of point sources, such growth curves are typically used to determine the most appropriate aperture radius for measurements, as the polarisation reaches an asymptotic value with increasing aperture \citep{bagnulo2011,bagnulo2016}. Variations with aperture size are likely due to the passing of background stars whose flux contaminate measurements. For extended objects, however, such variations may be indicative of properties intrinsic to the object, reflecting a dust environment which varies with photocentric distance. These curves will be particularly useful when analysing the polarimetric maps. 

Table \ref{tab:obslog} reports the numerical values measured from aperture polarimetry. The VLT values have been integrated in an aperture radius equivalent to 100 km (0.1 on the horizontal axis of Figure \ref{fig:growthcurves}). The recorded NOT measurements are the weighted averages of the repeated observations (two to four times in each filter) of each night. For smaller $\Delta$ values ($0.07-0.09$ AU), we chose aperture radii corresponding to 100 km as we did with our VLT measurements. When $\Delta$ is larger, aperture radii of 100 km corresponded to a pixel radii smaller than 5 pixels. We, therefore, opted for larger radii to assure sufficient flux in the aperture. The flat polarimetric growth curves from 2022-0ct-25 forward assure us that a choice of a larger aperture radius in the NOT observations should not change the value in polarisation. For completeness, we have included the spectropolarimetric measurements by  \cite{bagnulo2023} in the table, indicated with an asterix (*VLT). These observations use a rectangular aperture fixed to a 2" slit width, thus, covering a larger area than the imaging polarimetric measurements. 

\begin{figure*}
\begin{center}
    \centering
    \includegraphics[width=15cm]{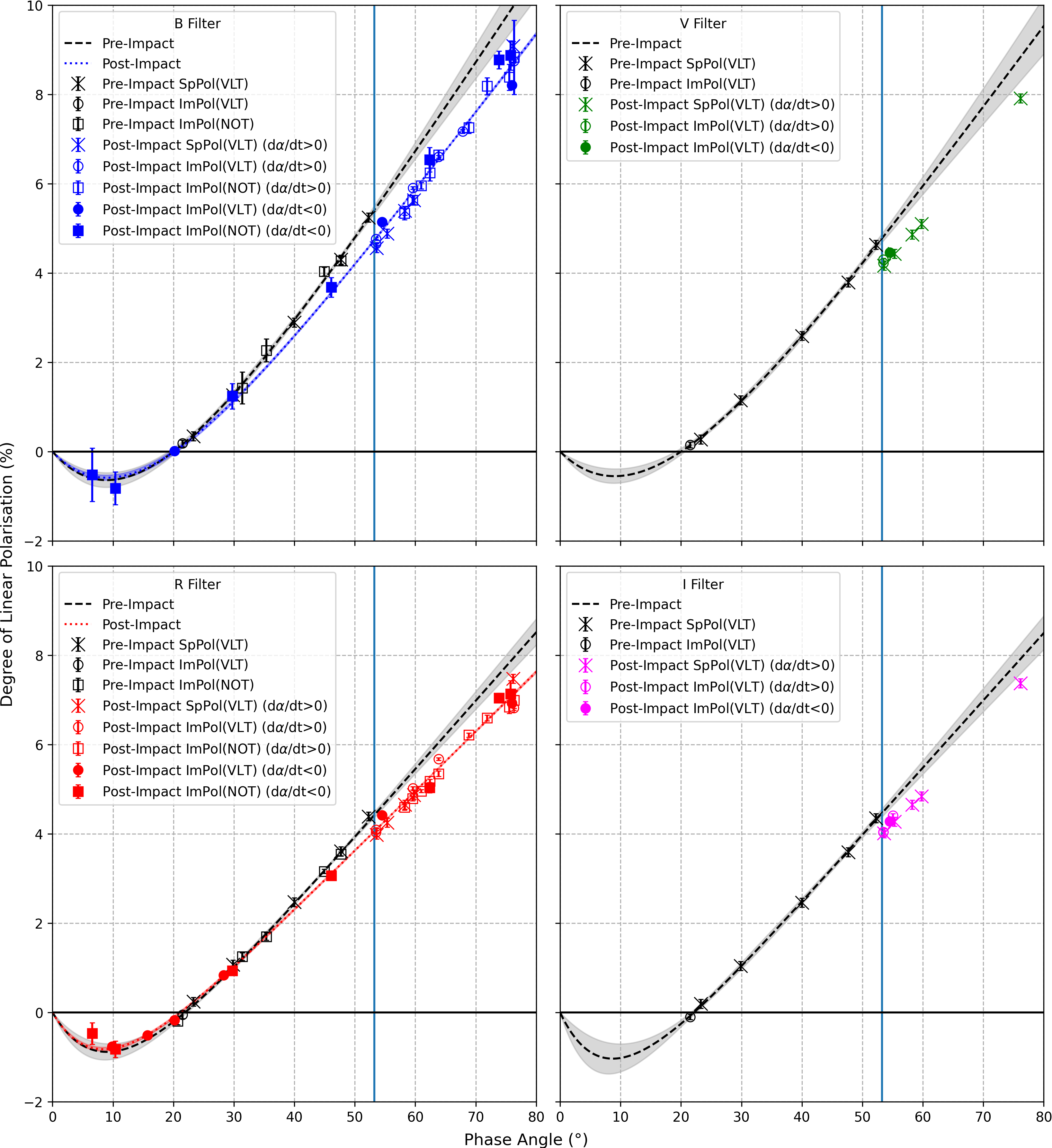}
    \caption{The polarimetric phase curves of Didymos-Dimorphos in the BVRI filters. The vertical blue line indicates the phase angle of the asteroid system at the time of the DART impact ($\sim 53.2^{\circ}$). The SpPol data points refer to the spectropolarimetric measurements taken with the VLT by \cite{bagnulo2023}, while the ImPol data points refer to the imaging (aperture) polarimetric measurements taken with the VLT and NOT (this study). In each plot, the black symbols and curve represent pre-impact measurements and their best fit, while the coloured symbols and curve represent post-impact measurements and fit. The shaded area around the curves (where applicable) represents $1\sigma$ of the fit. In the case of the ImPol measurements, the open and filled data points are measurements taken when the phase angle was increasing ($\rm{d}\alpha/\rm{d}t > 0$) and decreasing ($\rm{d}\alpha/\rm{d}t < 0$, after $\alpha$ peak at $76.34^{\circ}$ on 2022-Oct-21), respectively.}
    \label{fig:phasecurve}
\end{center}
\end{figure*}

\begin{figure*}
\begin{center}
    \centering
    \includegraphics[width=13cm]{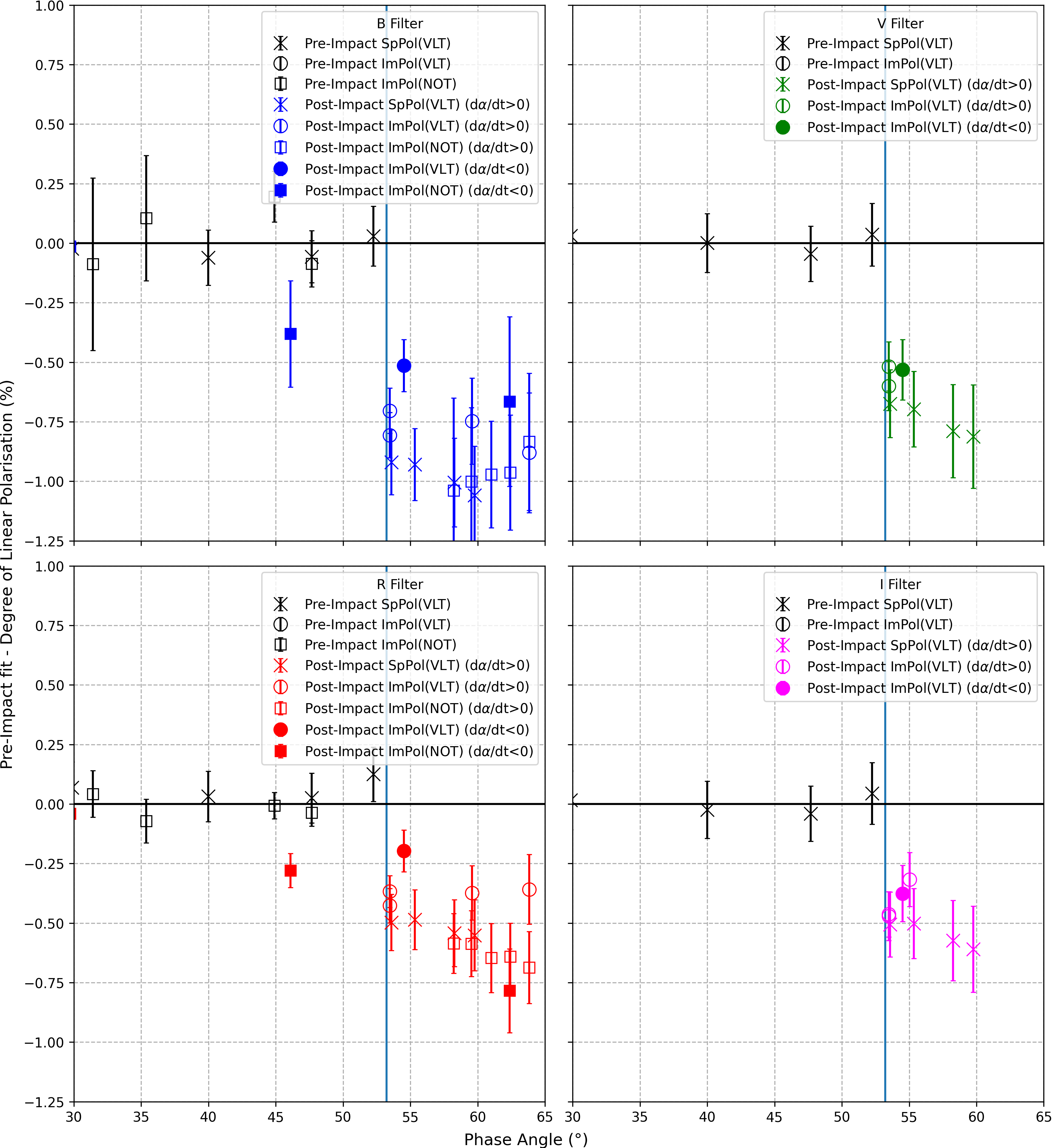}
    \caption{The difference between the calculated pre-impact curve and the measurements of the degree of linear polarisation, i.e., the value of the pre-impact curve minus the value of data points between phase angles $30-65^{\circ}$. The symbols are as those in Figure \ref{fig:phasecurve}.}
    \label{fig:phasecurvediff}
\end{center}
\end{figure*}

%

\begin{table*}
\caption{\label{tab:phasecurve} Numerical values according to the best fit of the pre- and post-impact fits of the polarimetric phase curve. $P(\alpha_{\rm{Impact}}$) is the value of polarisation at the impact phase angle, $53.2^{\circ}$, while $\Delta P(\alpha_{\rm{Impact}}$) is the difference between the pre- and post-impact values, \ainv is the inversion angle, \textit{h} is the slope of the curve at \ainv and $\rho_v$ is the geometric albedo calculated from Eq. \ref{eq:albedo}.}
\begin{center}
\begin{tabular}{rr@{$\pm$}lr@{$\pm$}lr@{$\pm$}lr@{$\pm$}l}
\hline\hline    

 & \multicolumn{2}{c}{Pre-Impact} & \multicolumn{2}{c}{Post-Impact} & \multicolumn{2}{c}{Pre-Impact} & \multicolumn{2}{c}{Post-Impact} \\ 
 \hline
 & \multicolumn{4}{c}{B} & \multicolumn{4}{c}{R} \\
$P(\alpha_{\rm{Impact}}$) (\%) & 5.40 & 0.09 & 4.73 & 0.02 & 4.42 & 0.06 & 4.05 & 0.01 \\
$\Delta P(\alpha_{\rm{Impact}}$) (\%) & \multicolumn{4}{c}{0.68 $\pm$ 0.09} & \multicolumn{4}{c}{0.37 $\pm$ 0.06} \\
\ainv ($^{\circ})$ & 20.0 & 0.1 & 20.2 & 0.1 & 21.9 & 0.1 & 21.1 & 0.1 \\
\textit{h} & 0.10 & 0.07 & 0.09 & 0.01 & 0.11 & 0.06 & 0.10 & 0.01 \\
\hline
 & \multicolumn{4}{c}{$\rm{V_B}$} & \multicolumn{4}{c}{$\rm{V_R}$} \\
 
$P(\alpha_{\rm{Impact}}$) (\%) & 4.86 & 0.09 & 4.26 & 0.02 &4.66 & 0.06 & 4.25 & 0.02 \\ 
$\Delta P(\alpha_{\rm{Impact}}$) (\%) & \multicolumn{4}{c}{0.61 $\pm$ 0.09} & \multicolumn{4}{c}{0.40 $\pm$ 0.06} \\
\ainv ($^{\circ})$ & 20.0 & 0.1 & 20.2 & 0.1 & 21.5 & 0.1 & 21.1 & 0.1 \\
\textit{h} & 0.09 & 0.08 & 0.08 & 0.01 & 0.12 & 0.05 & 0.11 & 0.01 \\
$\rho_v$ & 0.23 & 0.20 & 0.26 & 0.03 & 0.18 & 0.10 & 0.19 & 0.02 \\
\hline \hline
\end{tabular}
\end{center}
\end{table*}

\subsection{Polarimetric Phase Curve}
\label{sec:results_phasecurve}
Figure \ref{fig:phasecurve} shows the polarimetric phase curve of Didymos-Dimorphos in all four BVRI filters, with values obtained from aperture polarimetry plotted as a function phase angle, along with the results from \cite{bagnulo2023}. The pre-impact and post-impact data and fits are distinguished with black and coloured points and curves, respectively. Over the course of the entire observing run, the phase angle ranged between $\sim 7$ and $-76^{\circ}$. The phase angle peaked around half way through the observing run (2022-Oct-21) after which the phase angle decreased until reaching a minimum (2023-Jan-11). Consequently, several measurements were taken at very similar phase angles, but months apart, corresponding to the asteroid system's different stages with respect to the DART impact. We have distinguished these data points with open ($\rm{d}\alpha/\rm{d}t > 0$) and filled ($\rm{d}\alpha/\rm{d}t < 0$) data points for clarity. For example, around phase angle $52-54^{\circ}$, the system was observed immediately before (T-15hr), immediately after (T+5hr) and months after (T+58d) the impact. This extensive data set allows us to study the immediate as well as the long term effects of the DART impact. 

The best fits of pre- and post-impact data have been calculated with the linear-exponential empirical model suggested by \cite{muinonen2009}:
\begin{equation}
    P(\alpha) = A(\rm{e}^{-\frac{\alpha}{B}} - 1) + C\alpha,
    \label{eq:cellinofunc}
\end{equation}
where $\alpha$ is the phase angle, and $A, B, C$ are free parameters which shape the curve and are derived by least-square fitting. In the case of the V and I filters, there was insufficient data to calculate the post-impact best fit. To better study the differences between the pre- and post-impact polarimetric phase curves (of R and B filters only), we calculated the values of the best fits at the impact phase angle ($53.2^{\circ}$), the inversion angles $\ainv$, and the slope of the curves at $\ainv$ in Table \ref{tab:phasecurve}. In the context of asteroid polarimetry, Umov's law describes the empirical inverse correlation between the polarisation and the geometric albedo of the scattering surface---darker surfaces polarise light to a higher degree than brighter surface. A consequence of the Umov effect is the direct relation between the slope \textit{h} of the polarimetric phase curve at the inversion angle \ainv to the geometric albedo $\rho_v$. \cite{zellner1976} showed that the polarisation phase curve can be used to derive the albedo using two empirical relations:
\begin{equation}
\label{eq:albedo}
\begin{split}
    \log(\rho_v) &= C_1 \log(h) + C_2, \\
    \log(\rho_v) &= C_3 \log(\pmin) + C_4,
\end{split}
\end{equation}
where $C_1, C_2, C_3, C_4$ are constants that may be obtained from fitting. We used values $C_1 = -1.111 \pm 0.031$ and $C_2 = -1.781 \pm 0.025$ as suggested by \cite{cellino2015} to calculate the geometric albedo in this study. These constants, however, have been calculated and calibrated according to V-band polarimetric data. Therefore, we first converted our R- and B-band measurements to the V-band before using the above expression to calculate the geometric albedo. According to the spectropolarimetric measurements of \cite{bagnulo2023}, the ratio between measurements in the V- and R-bands are nearly constant with phase angle, i.e. $(\pq(\rm{V})-\pq(\rm{R}))/\pq(\rm{R}) \sim 0.05$, as well as between the V- and B-bands,  $(\pq(\rm{V})-\pq(\rm{B}))/\pq(\rm{B}) \sim 0.1$. Knowing this, we approximate the polarisation in the V filter using
\begin{equation}
\label{eq:r2v}
\begin{split}
    \pq(V_R) &= 0.05\pq(R) + \pq(R)\ , \\
    \pq(V_B) &= 0.1\pq(B) + \pq(B)\ .
\end{split}
\end{equation}
In Table \ref{tab:phasecurve}, we have recorded the properties of the $\rm{V_R}$ and $\rm{V_B}$ polarimetric phase curves, as well as our calculated values of the geometric albedo.
We measured a sharp drop in polarisation immediately after the impact. According to the best fit curves, this drop was approximately $0.7$ p.p. and $0.4$ p.p. in the B and R filters, respectively. In the following days and weeks the polarisation increased with phase angle as expected, but remained lower than the extrapolated pre-impact polarisation. We observed some scattering in values around the peak phase angle, the reason for which is still unclear. This scattering appears more pronounced in the B filter compared to the R filter due to the generally higher value of polarisation in the blue. After the peak phase angle ($76^{\circ})$, the polarisation decreased as the phase angle decreased, again remaining at a lower level compared to the pre-impact system. In particular, the data points around phase angle $52-54^{\circ}$ are of particular interest. In Figure \ref{fig:phasecurvediff}, we plotted the difference between the fit of the pre-impact data and the observed polarisation in the BVRI filters around phase angles $30-65^{\circ}$. We focus on these phase angles only due to the unreliability of the pre-impact fit at larger and smaller phase angles. This plot highlights that the polarisation appears to remain lower than the pre-impact level even months after the impact. The similarity of the pre- and post-impact data points at small phase angles (approaching $20^{\circ}$) is due to the viewing geometry of the polarimetric phase curve. We cannot compare the negative polarisation branch of the pre- and post-impact data sets due to lack of data. 

Overall, our measurements are in good agreement with spectropolarimetric measurements by \cite{bagnulo2023}. The observations in that study, however, only cover the pre-impact and early post-impact stages of the DART impact. The new data presented here is specially suitable to establish whether, months after the impact, the polarisation characteristics of Didymos-Dimorphos came back to the pre-impact situation or if they were still affected by the dust ejecta. We conclude that the change in polarisation induced by DART lasted long after the impact, when the majority of the initial dust ejecta is expected by most to have dispersed. We explore some explanations for this finding in the Discussion section.

\subsection{Imaging Maps}
\begin{figure*}
\begin{center}
    \centering
    \includegraphics[width=\linewidth]{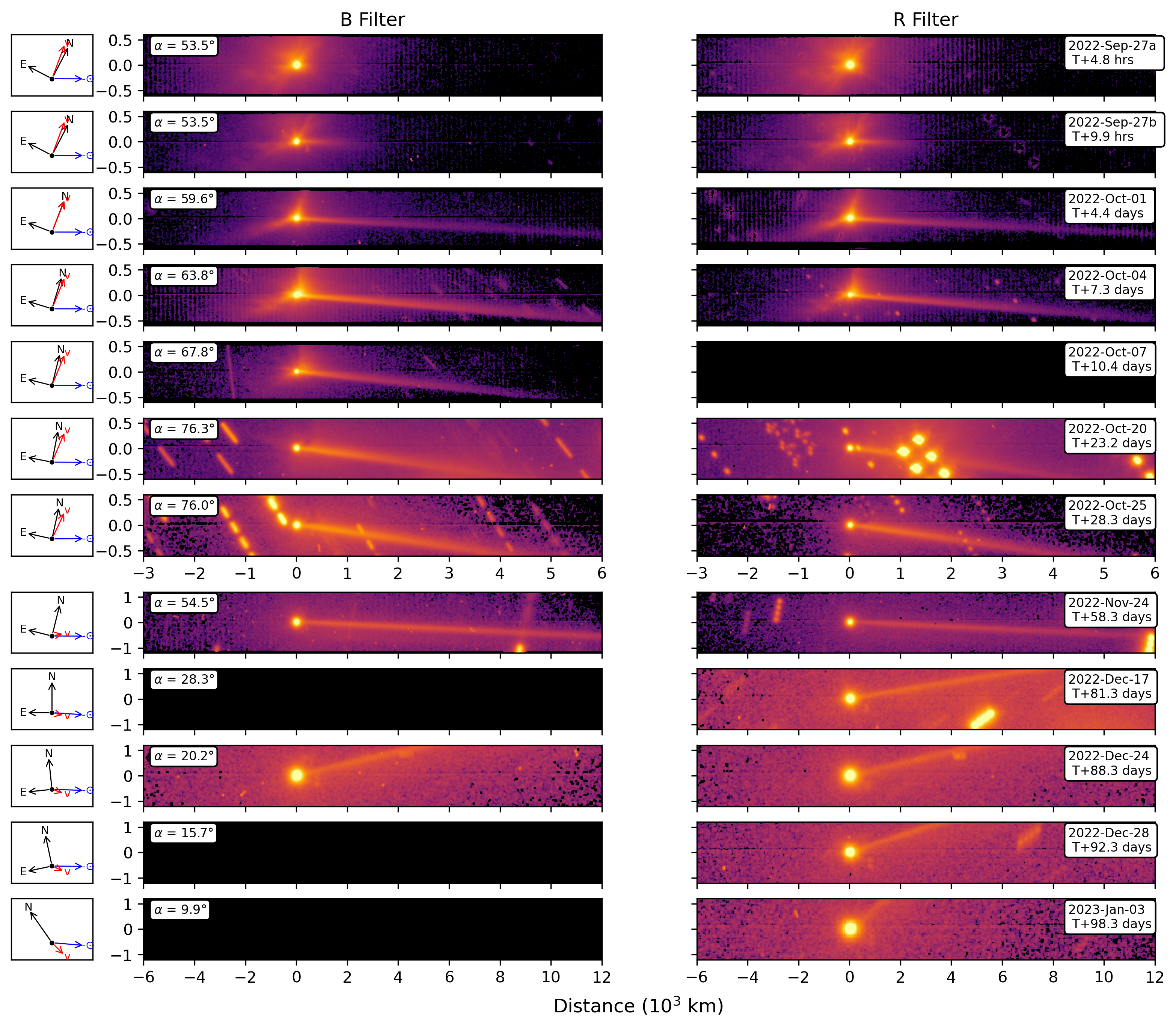}
    \caption{Imaging maps of VLT data of Didymos-Dimorphos in the B and R filters. The phase angle and date of each observation are indicated in each row. The north (N), east (E), anti-solar ($-\odot$) and velocity ($v$) directions are given in the panels to the left of the maps. Panels are left blank (black) in cases where there is no data. Background stars appear as aligned and sometimes elongated bright spots (e.g. 2022-Oct-20, B filter).}
    \label{fig:deeps}
\end{center}
\end{figure*}

Figure \ref{fig:deeps} displays the imaging maps in the B- and R-filters created from the VLT data. Each row represents maps from a given epoch and phase angle, shown in the text boxes. The panels to the left of the maps indicate the north (N), east (E), anti-solar ($-\odot$) and velocity ($v$) directions. The bottom five panels are plotted with a distance scale double to that of the upper seven panels. The maps have been plotted in a logarithmic scale for display purposes and those that appear blank (black) are due to the absence of data in a given filter on a given date. The presence of background stars appear as bright, sometimes elongated, spots in the maps.  

Our observations reveal effects of the DART impact consistent with findings from HST \citep{li2023} and MUSE \citep{opitom2023} observations. At T+4.8hr, the ejected material forms a cone-shaped cloud in the eastern hemisphere that engulfs the entire asteroid system. Several linear features are apparent in these initial images and become more defined in subsequent observations. The most pronounced of these features, found in the northern and south-eastern directions, are the edges of the ejecta cone. Similarly, faint ``clumps" of material are visible moving radially away from the asteroid. At T+9.9hr, the first signs of tail formation appear, extending up to 1500 km in length. By T+4.4d, the tail had stretched out beyond 5000 km and continued to grow with time, eventually extending beyond 10 000 km as more material was pushed towards the anti-solar direction by solar radiation pressure. 

At T+7.3d and T+10.4d, faint linear features in the direction parallel to the tail appear within the dust cloud. These features likely arise from the natural separation of particles of different size caused by radiation pressure, where smaller particles are accelerated more efficiently than larger particles \citep{li2023}. By T+23.3d, most of the material forming the dust cloud has dissipated, while the tail remains. From this point, little variation is found except for the gradual attenuation of the intensity of the tail. 

\begin{figure*}
\begin{center}
    \centering
    \includegraphics[width=\linewidth]{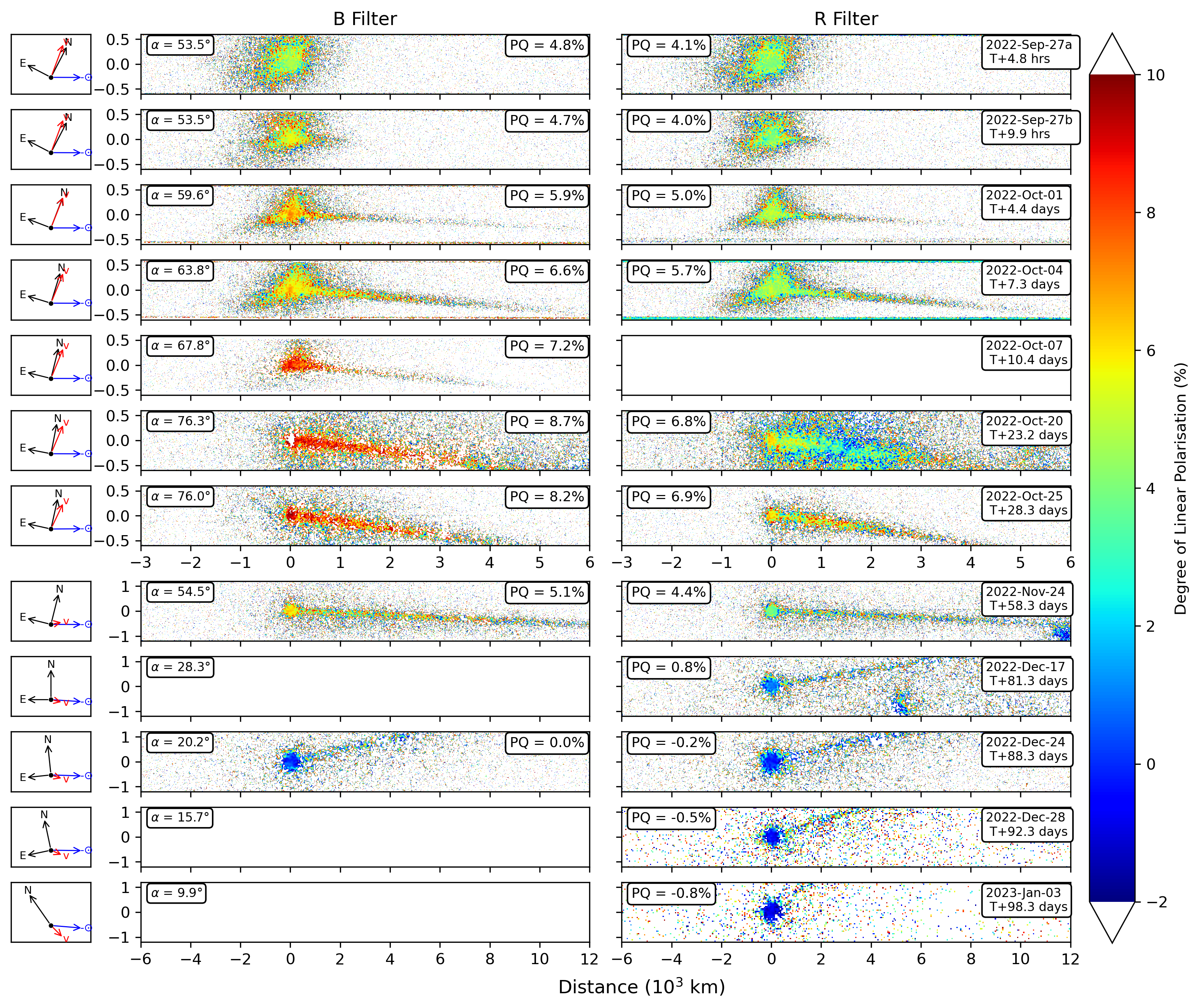}
    \caption{Polarimetric maps of VLT data of Didymos-Dimorphos in the B and R filters. The colour of each pixel represents the value of polarisation as given in the colourbar. Pixels with values outside the $(-2, 10)$ \% range have been set to white. The phase angle, date of each observation, and value measured from aperture polarimetry (\pq) are indicated in each row. The north (N), east (E), anti-solar ($-\odot$) and velocity ($v$) directions are given in the panels to the left of the maps. Panels are left blank in cases where there is no data. Gaps in the tail and the spurious tail of 2022-Oct-20 (R filter) are due to background stars.}
    \label{fig:polmaps}
\end{center}
\end{figure*}

\begin{figure*}
    \centering
     \subfigure[]{
        \includegraphics[width=0.9\linewidth]{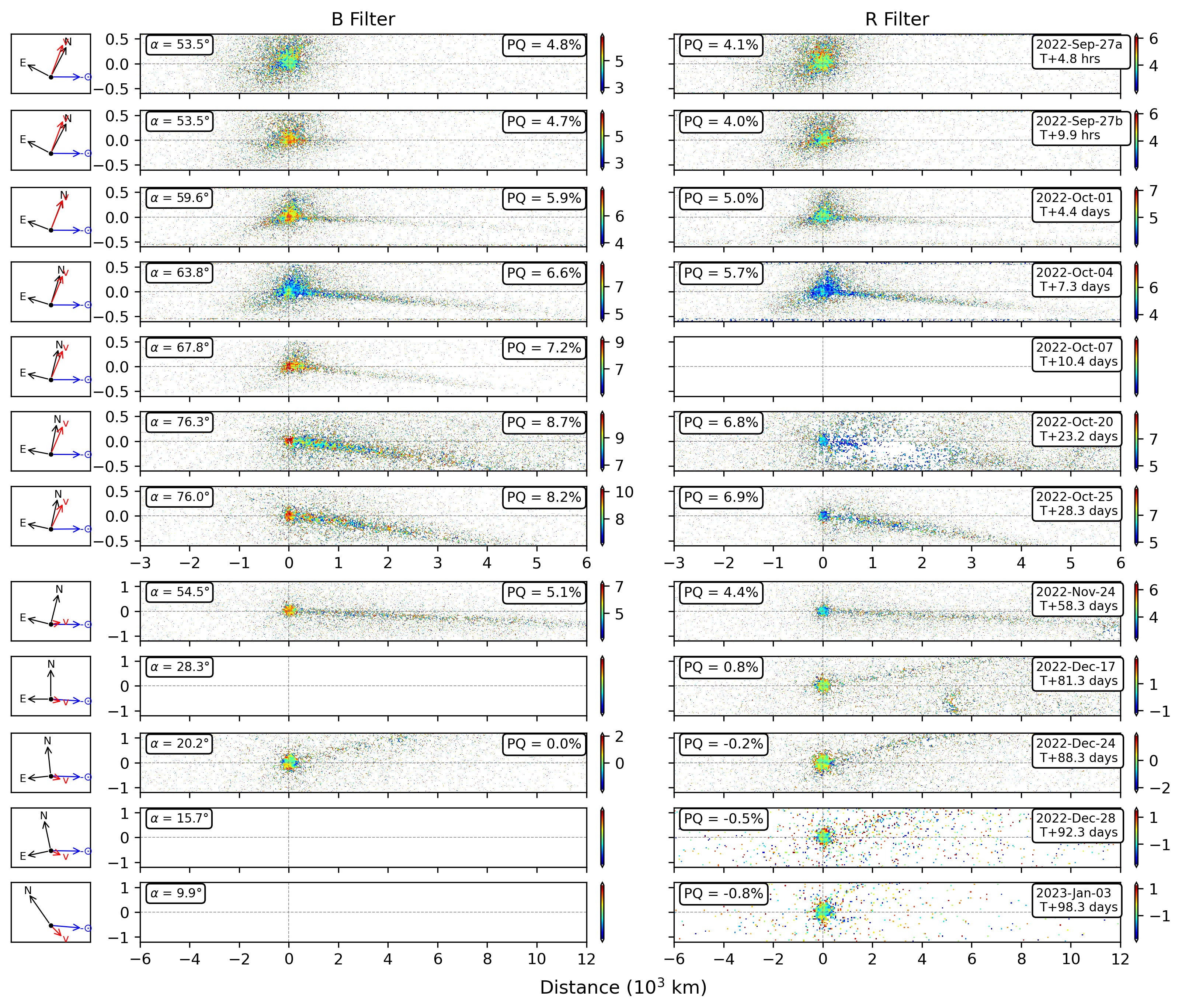}
        \label{fig:polmaps_scaled}}
     \subfigure[]{
        \includegraphics[width=0.7\linewidth]{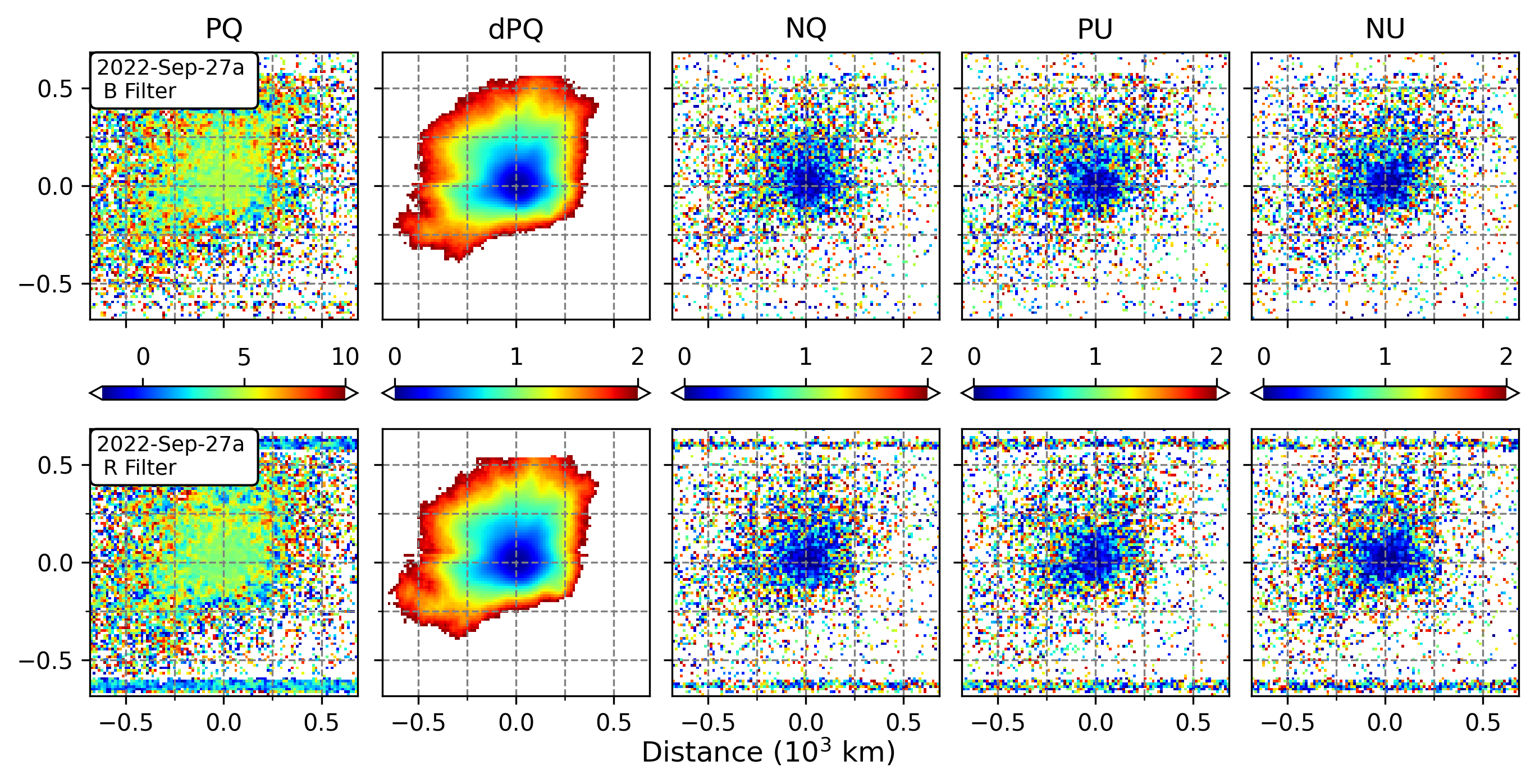}
        \label{fig:polmaps_QC}}
    \caption{(a) The same polarimetric maps as in Figure \ref{fig:polmaps}, but with dynamic colour scales. The scale of each panel has been set to \pq\ $\pm 2$ \%, where \pq\ is the value measured from aperture polarimetry and is indicated in each panel. The grid indicates the photocentre of the asteroid system. (b) Quality checks of 2022-09-27a polarimetric maps. From left to right: \pq\ map with $(-2, 10) \%$ scale, $dP_Q$  map, \nnq\ (Null) map, \pu\ map, \nnu\ map. The latter three maps are plotted with absolute values. A colourbar is given for each column.}
    \end{figure*}



\subsection{Polarimetric Maps}
Figure \ref{fig:polmaps} displays the polarimetric maps in the B- and R-filters created from VLT data. Again, each row represents an observing epoch, with directional labels (north, east, anti-sunward, and velocity vector) in the panels to the left of the maps. The colour of each pixel represents the value of polarisation as given in the colourbar to the right of the plots. The scale of the colourbar has been set to the same scale as the polarimetric phase curve (dark blue at $-2\%$ to dark red at $10\%$). For viewing purposes, all pixels whose value lays outside this range has been set to white. The polarisation value measured from aperture polarimetry is indicated by \pq\ in a text box of each plot. The same polarimetric maps are also presented in Figure \ref{fig:polmaps_scaled}, but this time the scale has been set to \pq\ $\pm 2 \%$ in each plot. For example, since \pq\ $= 5 \%$ in the R-filter on 2022-Oct-01, the maximum value in this plot will be $7 \%$ (red) and the minimum will be $3 \%$ (blue), while pixels with values outside this range appear in white. The reason for plotting the polarimetric maps in this manner is to increase the contrast and enhance any variability in polarisation. A grid has been included in these maps to indicate the photocentre. Same as before, the maps that appear blank are due to absence of data and the presence of background stars create some spurious signals. Figure \ref{fig:polmaps_QC} shows an example of some of the quality checks we have performed for our polarimetric maps. In particular, they show the uncertainty map of \pq\, i.e. $dP_Q$, as well as the \nnq\ (Null), \pu\, and \nnu\ maps, using data from 2022-Sep-27a in both the B- and R-filters as an example. The latter three maps have been plotted in absolute values. The scale of each panel corresponds to the colourbar in the centre of its column. As expected, the \pu\ and Null maps oscillate around zero, while the uncertainty of \pq\ increases significantly with photocentre distance. Although not displayed, we assured the accuracy of our other polarimetric maps with such quality checks. 

The effects of the DART impact are apparent in polarised light. The material ejected immediately after the impact polarises the light in an almost uniform manner, with some small scale variations (i.e., fluctuations of red and blue) at the edges where the dust cloud tapers. Despite this, these variations do not appear as a clear gradient such as those in \cite{rosenbush2017}, so we attribute these small scale variations to noise. This homogeneity is reflected in the generally flat polarimetric growth curves of the first epochs in Figure \ref{fig:growthcurves}. Compared to the imaging maps, the strong linear features found in the northern and south-eastern directions (cone ``edges") are not as clearly defined in the polarimetric maps. Instead, the material filling the entire cone polarises light almost uniformly. This may be due to the fact that the degree of linear polarisation is a relative quantity, which, in contrast to brightness, does not depend on the number of particles but rather the physical properties (size, shape, structure, composition). The tail is first visible at T+9.9hr in polarised light, appearing as an extension of the dust cloud. As it extends with time, it appears shorter in polarised light compared to the imaging maps, again, due to diminishing brightness of the tail with distance. A further consequence is the difficulty of measuring any variability of polarisation within the tail. The spurious thick tail in R of 2022-Oct-20 is due to the passing of background stars.

Despite the overall homogeneity of polarisation of the dust cloud, a feature is visible at the photocentre in many of the polarimetric maps. This feature is more obvious in Figure \ref{fig:polmaps_scaled} where the degree of polarisation of the photocentre is $\sim 0.5-1 \%$ higher than the surrounding dust cloud. The consistent alignment with the photocentre and its point-like appearance lead us to believe that this features represents the asteroid surface. In other words, the dust cloud is somewhat transparent to the asteroid surface in some epochs and is opaque in others. As the majority of the dust cloud disperses (around T+23.2d), a $\sim 1 \%$ difference in polarisation persists between the photocentre and the tail (except for the later epochs where the tail is too faint to distinguish a gradient). This gradient suggests that the properties of material on the surface of the asteroid(s) differ from those of the material in the dust cloud and tail. These variations are also reflected in the polarimetric growth curves of these epochs, where the polarisation peaks at small apertures and gradually decreases with increasing aperture. As the aperture size increases, more of the signal from the lower-polarised cloud and tail is measured, bringing down the overall measurement of average polarisation within the aperture. Interestingly, the asteroid surface is visible in the B filter at just T+9.9hr, while it does not become visible in the R filter until T+7.3d. This observation may have implications for understanding the temporal evolution of the size distribution of the material within the dust cloud. Finally, in the case that this interpretation is correct, there is a possibility that the polarisation of the asteroid surface is, in fact, higher than what we find in our measurements. Given that the asteroid is sub-pixel and unresolved, its image becomes dispersed and diluted with the surrounding cloud of particles. Thus, if the signal is contaminated by the depolarising dust cloud, the polarisation of the asteroid surface could exceed our measurements. This is difficult to confirm, however, without data of higher spatial resolution.


\section{Discussion}\label{Sect_Discussion}

\subsection{Drop in Polarisation I: Immediate Effects of DART}
The most obvious effect of the DART impact is the dramatic drop in polarisation immediately after impact. Based on this, we can safely assume that particles different to those present on the original surface(s) (pre-impact) are injected into the system. Importantly, they have a neutralising effect on polarisation. For the interpretation of this results we turn to light scattering theory, in particular, studies involving laboratory experiments and computer simulations. In this context, studies can deal with various particle size ($r$) to wavelength ($\lambda$) ratio. The Rayleigh regime is valid for particles much smaller than wavelength ($r \ll \lambda$), the resonance regime for particles of similar size to wavelength ($r \sim \lambda$), and the geometric-optics regime for particles much larger than wavelength  ($r \gg \lambda$). The rules of light scattering vary according to these size regimes. 

Several studies have shown that the value of maximum polarisation (\pmax) and its position on the phase curve (\amax) depend strongly on the scattering particle size. In the Rayleigh-resonance regime ($r \ll \lambda$ or $r \sim \lambda$), \pmax tends to increase as particle size decreases. In the geometric optics regime ($r \gg \lambda$), however, this rule is reversed: \pmax increases with increasing particle size. As the size of the particles increases into the geometric optics regime, \amax shifts towards larger phase angles. This correlation is found, for example, in the laboratory experiments by \cite{munoz2021} who studied the light scattering effects of particles over the full size domain and found that \pmax is approximately 20\% higher and shifts towards a larger phase angle for millimeter-sized pebbles compared to a similar sample with particles in the $100 \mu$m range. Further, the size distribution of the overall sample affects the polarimetric properties. Both \cite{escobar2018} and \cite{frattin2022} found that \pmax and \amax increased, while the negative polarisation branch (NPB) almost disappears, when they sifted their sample to remove small particles (up to tens of $\mu$m in radius). In other words, in the geometric optical regime, it appears that the fraction of small particles included in the overall sample with various particle sizes determines the maximum degree of polarisation. This is in agreement with the findings of light scattering simulations using Gaussian random sphere shapes by \cite{liu2015}, who also concluded that small particles limit the maximum degree of polarisation as the mean size of the particles moves into geometric optics range.

Likewise, the colour of the scattering material affects the polarimetric behaviour of the system. As noted by \cite{umov1905}, darker surfaces polarise light to a higher degree than brighter surfaces. Therefore, the maximum degree of polarisation is dependent not only on the size of the particles but on their refractive index. Laboratory measurements of low, moderate to highly absorbing millimeter-sized compact particles reveal a direct correlation between the imaginary part of the refractive index of the sample and the maximum degree of polarisation \citep{munoz2020}. \pmax is significantly higher for highly absorbing particle (low albedo) than for the low absorbing particle (high albedo), while \pmax of the moderately absorbing particle lies within the range of the other two samples. This may be interpreted as being the result of multiple scattering, which randomises the polarisation plane at each scattering and, consequently, neutralises the polarising effects of single scattering \citep{sorensen2001}.

The situation becomes more intricate when multiple variables are considered simultaneously---disentangling their individual contributions and/or identifying the dominant influencing factor (size, structure, compositions, etc.) on the polarimetric properties is challenging. For example, in a study using materials of various size distributions and compositions (hence, colour), \cite{frattin2019} found that the Allende meteorite sample presented the lowest \pmax value, despite its dark colour, lower than even the samples whitest in colour. This result was attributed to the Allende sample having the largest concentration of small particles, thus, reducing the value of maximum polarisation. Further, a principal difficulty that arises in the interpretation of astronomical observations with laboratory and computational studies is the non-uniqueness of light scattering properties of different particle groups. This is because the light scattering properties of various particle types does not change drastically between the different types. For instance, under single-scattering conditions, both small particles in the Rayleigh regime ($r \ll \lambda$, where $r$ is the particle size) and large particles in the geometric optics regime ($r \gg \lambda$) polarise light to a high degree. The position of \pmax for Rayleigh particles is $90^{\circ}$, whereas it is shifted towards larger phase angles for $r \gg \lambda$. 

Based on the conclusions from previous studies mentioned above, a possible, but not exclusive, interpretation is that the ejected particles are smaller and/or brighter than those present on the pre-impact surface. The continuing discussion is based on this interpretation. High resolution images returned by the camera onboard the DART spacecraft, Didymos Reconnaissance and Asteroid Camera for Optical Navigation (DRACO), provided insight into the geological composition of the pre-impact surface of Dimorphos. The imagery revealed that the terrain consisted of sizable boulders in the order of tens of meters, as well as smaller cobbles and pebbles \citep{daly2023,barnouin2023}. A possible explanation is that that the impact of the DART spacecraft caused the fragmentation of many of these larger elements and/or the ejection of material from under the upper-most surface of Dimorphos. Identifying the exact mechanism responsible for the change in particle size distribution, however, is beyond the scope of this study. Although the phase angle range of our measurements does not cover the maximum degree of polarisation, we can infer from our measurements that \pmax of the pre-impact system is larger than that of the post-impact system. From this, we can conclude that the drop in polarisation is (partly) due to the ejection of particles smaller than those present on the original surface, although large enough to scattering light in the geometric-optics regime. Furthermore, the slope of the post-impact curve at the inversion angle is smaller than that of the pre-impact curve. The smaller slope, and hence higher albedo, of the post-impact data compared to the pre-impact data implies that the material ejected by the impact is brighter than that of the original surface. We understand this to be due to the effects of space weathering, which refers to the gradual alteration of exposed surfaces via their interaction with the space environment. This exposure encompasses various sources of energetic radiation (e.g., solar, galactic, and magnetospheric, where applicable) and meteoroid-like objects which can strike the surface. In the case of atmosphereless objects, one of the consequences of space weathering is believed to be the darkening of the surface \citep[e.g.]{hendrix2019,hasegawa2022,zhang2022,escobar2018b,penttila2020}. This outer space-weathered crust may act as a protective layer to the material below, conserving its more-pristine nature. In this context, it is reasonable to assume that the material below the surface of Dimorphos is brighter, and lower in polarisation, than the outer dark layer. The closest and most obvious place where the effects of space weathering is visible is the Moon. As far back as 1955, \cite{gold1955} recognised that material in and around impact craters, including long ejecta streaks strewn across the surface, are brighter than the surrounding material. Dynamic modelling of the cratering process show that much of the material ejected by the DART impact comes from the inner layers of the asteroid \citep{ferrari2022}.  Although not spatially resolvable, it is possible that the DART impact ejecta left a footprint on the surface of Didymos, while the surface of Dimorphos may be completely refreshed due to the loss of a portion of the original material after impact. The removal of a space-weathered surface layer is expected to rejuvenate the surface in such a way as to increase its albedo, thus decreasing the polarisation.

\subsection{Drop in Polarisation II: Long-term Effects of DART}
Our monitoring campaign allows us to address whether the drop in polarisation is limited to the immediate aftermath of the impact or if it persists as a long-term phenomenon. After the Deep Impact mission for example, the polarisation of Comet 9P/Tempel 1 returned to its pre-impact level within a few days \citep{harrington2007,hadamcik2007,furusho2007}. The polarimetric maps of Didymos-Dimorphos show that the system was engulfed by the ejecta dust cloud in the early weeks after the impact. The aperture polarimetric measurements taken around this time are obviously influenced by the still-present dust cloud made of smaller and/or brighter material, hence, bringing down the value of polarisation. The maps at around T+23.2d and T+28.3d, however, show that the majority of the dust cloud had dissipated and, thus, observations from this point forward should be less influenced by the ejecta. As discussed before, these dates also happen to coincide with the maximum phase angle as viewed from Earth during this period ($\sim 76^{\circ}$). The decreasing phase angle allows us to measure and compare the polarisation of the system at similar phase angles but different stages in the context of the DART impact. 

Figure \ref{fig:phasecurve} shows that the polarisation of the system in this long-term post-impact stage, after the dissipation of the majority of the dust cloud, continues to exhibit a lower level compared to the pre-impact state. However, this difference may not be as drastic as observed immediately after the impact. We are considering two potential explanations for this phenomenon. Firstly, it's possible that some remnants of the ejecta cloud, not currently resolved by the VLT, remain in orbit within the system. This is highly likely considering the presence of the dust tail --- sustaining the tail for such a long time would be impossible without large amounts of material remaining in the system after the impact. Secondly, it's possible that some of the fresh ejecta has accumulated on the surfaces of Didymos and Dimorphos. In either case, the presence of remaining fresh material within the system could contribute to maintaining the lower level of polarisation. Depending on the properties of the remaining material, it could also affect the magnitude of the system. \cite{graykowski2023} found that Didymos-Dimorphos faded back to its 'normal' brightness (Hv = 18.1) around 23 days post-impact. Comparing to pre-impact measurements taken in the 2003 apparition, Pravec et al. (2023, private communication) measure a marginal change (brightening) of the mean (rotationally corrected) absolute magnitude H of just --0.061 magnitude that they determined from their photometric measurements taken at solar phases from 6.6 to $28.3^{\circ}$ from 2022-Dec-17 to 2023-Jan-29. However, this brightening has been detected at only 1.9-sigma level, not statistically significant and could be attributed to an error or anomalous observation.

Using various methods to model the early collisional processes of material within the ejecta, \cite{ferrari2022} found that among all of the ejecta fragments, 55\% of particles are still in orbit after a week, 9\% reaccumulate on the surface of Dimorphos and 36\% escape the system. The study does not detail the reaccumulation of particles on the surface of Didymos and does not run beyond one week. With Monte Carlo models, \cite{moreno2023} provide a characterisation of ejection geometric properties and ejecta dust properties based on HST observations of Didymos-Dimorphos. They suspect that the generation of the secondary tail is associated to reimpacting material on the surface of Didymos. Their calculations show that up to $1.5 \times 10^6$ kg of material reimpact the surfaces of both Dimorphos and Didymos in the first 20 days following the impact, some of which may have settled permanently on the surface. Further, any material still in orbit around this time will reside in larger particles (but still smaller than the pre-impact surface, considering the lower polarisation) as the smallest particle components will have been removed by radiation pressure. Due to the small effective cross section, this remaining mass will have very little effect on the total brightness some weeks after the impact, hence in agreement with measurements by \cite{graykowski2023} and Pravec et al. (2023, private communication). These findings lend support to our hypothesis that residual material from the ejecta cloud still remain within the system at the time of our observations, either in orbit or deposited on the asteroid surfaces. 

\subsection{Short-term Variations in Polarisation}

\begin{figure*}
\begin{center}
    \centering
    \includegraphics[width=12cm]{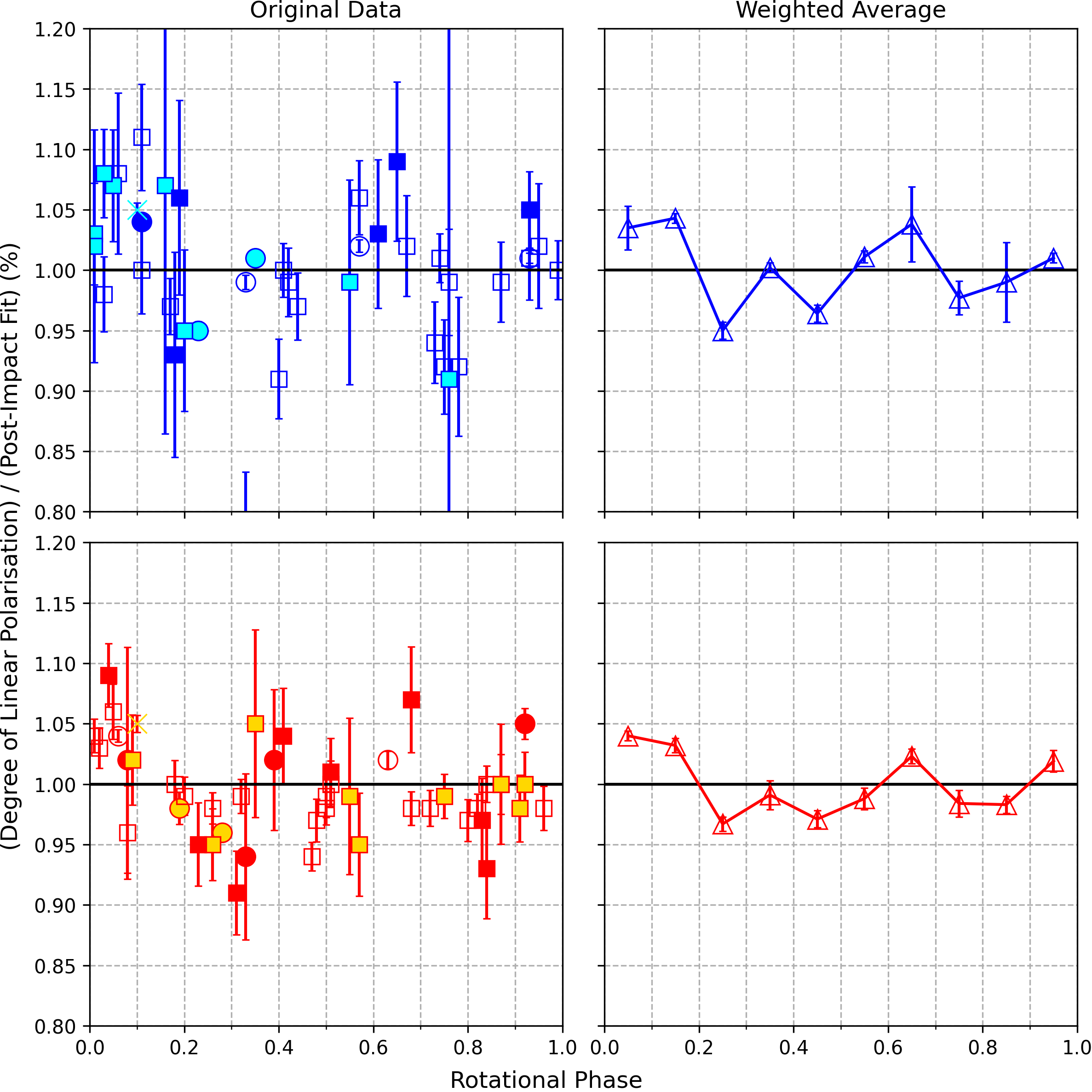}
    \caption{The ratio between measurements of polarisation of Didymos-Dimorphos (B- and R-filters) and the fit of the post-impact data at the corresponding phase angle, phased according to the rotation period of Didymos (2.26\,h). The left panels show the original, individual measurements compared to the fit, and the right panels show the weighted average of measurements grouped together in steps of 0.1 rotation fractions. The symbols are as in Figure \ref{fig:phasecurve}, with the data points around phase angles $75-76^{\circ}$ highlighted in light blue and yellow.}
    \label{fig:rotorb}
\end{center}
\end{figure*}

Across the polarimetric phase curve, some noticeable discrepancies are present among polarisation measurements obtained at similar phase angles. While some of these discrepancies are small, others exceed the estimated uncertainties and are apparent both in comparisons between measurements taken using different instruments and those obtained with the same instrument. These observed discrepancies have several potential implications. On one hand, they may indicate the presence of complex and variable physical processes occurring in the aftermath of the impact event, such as evolution and expansion of the ejecta cloud, changes in surface properties, or dynamical changes (gravitational effects, solar radiation pressure, etc.) in the cloud. On the other hand, they may be due to instrumental/data reduction effects or the observing dynamics of the system. Here, we explore some of these possibilities. 

First, we consider differences between measurements taken with different instruments. As visible from the polarimetric growth curves (Figure \ref{fig:growthcurves}), measurements taken in different apertures can yield different results. To account for this, we systematically chose the pixel radii that best approximated 100 km for the VLT data. For NOT data, again, we chose the pixel radius closest to 100 km when possible. At larger asteroid-observer distances ($\Delta > 0.1$ AU), however, our measurements were taken in larger radii, as 100 km corresponded to pixel radii smaller than 5 pixels. Considering this only applies to pre-impact measurements, when the target is a point source, and post-impact measurements beyond 2022-Oct-25 ($\rm{d}\alpha/\rm{d}t > 0$), when most of the dust cloud had dispersed, a larger aperture size should not lead to significant differences compared to a 100 km aperture.  Lastly, the spectropolarimetric measurements cover a larger area than the imaging polarimetric measurements. Despite this, these measurements are consistent with our measurements within uncertainties in some cases, but not in others. 

Second, we consider variations which may be intrinsic to the asteroid system. On the night of the impact, two imaging- and one spectropolarimetric measurement were taken with the VLT and exhibit considerable discrepancies. These variations may be attributed to the dust environment at the moment of the observations, which was likely rapidly evolving immediately after the impact. Such rapid changes may have been mostly prevalent in the first days following the impact. More puzzling are the discrepancies observed around phase angle $75-76^{\circ}$, obtained more than 2 weeks after the impact, which amount to $\sim 0.8-0.9$ p.p. in the B and R filters.

Finally, we consider the possibility of the polarisation being modulated by the rotation of Didymos or the eclipsing events of Didymos and Dimorphos. Subtle but persistent spectral variability has been found in pre-impact observations by \cite{ieva2022} and De León et al. (2023, private communication, preliminary result) which appear to be in sync with the rotation of Didymos, hinting to compositional differences throughout the surface. No post-impact variability has been reported thus far. Although we do not have enough pre-impact data to search for rotational variability, we tested it with our post-impact data. To do this, we phased the post-impact data according to the rotation period of Didymos, 2.26\,h \citep{pravec2006}, and the new orbital period of Dimorphos, 11.5\,h \citep{daly2023}, and searched for a periodic variation with respect to the values predicted by the global fit to all data points. The two to four observation repetitions in each filter carried out on each observation night at the NOT proved especially useful for this test. We excluded the data taken in the earliest days after the impact but included the data from 2022-Oct-04 (T+7.3d) onwards, when the asteroid surface is visible through the dust cloud according to the polarimetric maps (Figure \ref{fig:polmaps_scaled}). We recognise that measurements taken around this time may be influenced by the dust cloud, and hence, any changes in the dust environment may lead to variations in the polarisation. However, we have chosen to include them considering the small aperture sizes of the measurements. 

In the left panels of Figure \ref{fig:rotorb}, we have plotted the ratio between each individual measurement of polarisation and the fit of the post-impact data (B- and R-filters) at that given phase angle as a function of the rotational phase of Didymos, where one full rotation corresponds to 2.26\,h and the each 0.1 interval on the horizontal axis is approximately 13 minutes. We have plotted the data according the mid-point of the observation and have plotted the multiple NOT measurements of each night separately (as opposed to the weighted average). We have highlighted the data points taken around phase angles $75-76^{\circ}$, where the largest variations are found, with light blue and yellow. In the right panels, we have plotted the weighted average of data grouped together in 0.1 rotation fraction intervals. Considering the matching trend in both the B- and R-filters, this plot shows marginal evidence of variation of polarisation with rotation of Didymos. Further investigations and observations are necessary to confirm this potential finding. If true, an open question remains as to whether this variation is linked to compositional differences of the pre-impact surface which have survived the impact, or due to the potentially non-uniform settling of fresh ejecta on the surface of the asteroid. We conducted the same test, phasing the data according to the orbital period of Dimorphos (11.5\,h), but did not find any correlation with the discrepancies.

Overall, it is difficult to establish the exact reason for the discrepancies in polarisation measurements. While some small discrepancies may be attributed to instrumental or observational effects, others may provide valuable insights into the nature of the ejected material, dust dynamics, and surface properties of the asteroid system. Further  investigations will be crucial in unraveling the underlying cause of these observed discrepancies and their implications. The future Hera spacecraft \citep{michel2018}, targeted to rendezvous with Didymos-Dimorphos in 2027 with the objective to study the asteroid system and the outcome of the DART impact, will provide the information that will clarify the situation.

\subsection{Polarisation of the Ejecta-Cloud and Tail}
Based on our polarimetric phase curve analysis, we conclude that the drop in polarisation is caused by the ejection of particles of material smaller and/or brighter than that on the original asteroid system. In a similar sense, one would expect to see variations in polarisation within the dust cloud, particularly in areas containing concentrations of particles of different properties. The chosen colour scale for our polarimetric maps indicates that regions with higher concentrations of larger particles would appear redder, while areas with smaller particles would appear bluer. Despite this, only minor features are discernible in these maps. Solar radiation pressure naturally separates particles of different sizes, with smaller particles generally being accelerated anti-sunward more efficiently than larger ones. We observe evidence of this effect in the imaging maps at T+7.3d and T+10.4d, where several linear features appear in the south-eastern ejecta cone ``edge" running parallel to the anti-solar direction. In the HST images, these features are visible from T+4.7d. Consequently, an inhomogeneous distribution of particles sizes should be present in this specific region of the cloud around these dates. Similarly, the northern edge of the ejecta cone expands to a wing-like shape with a sharp edge in the anti-solar direction, indicating a cut-off in the largest particle size of the ejecta \citep{li2023}. 

Signs of this inhomogeneous mix of particles are visible solely in the B filter polarimetric maps at T+4.4d (Figure \ref{fig:polmaps}). In particular, a clump-like feature, appearing redder in colour than the surrounding area, emerges in the south-eastern ejecta cone edge, implying a higher concentration of larger particles in that region. This agrees with the notion that particles are organised according to size, with smaller particles being blown in the anti-solar direction and leaving behind the larger particles. The sharp northern edge found in the imaging maps, however, is less evident in the polarimetric maps. Further, these features are not obvious in any of the other maps. The reason for this could be attributed to the S/N and/or the resolution of the polarimetric measurements. Alternatively, it may be a result of the mixing of particles of various sizes as these larger particles fall back towards the system. As discussed, the concentration of large vs. small particles within an overall sample has an effect on the polarimetric properties \citep{escobar2018b,frattin2022}.

As for the tail in the polarimetric maps, it appears as an extension of the dust cloud within the first ten days. In other words, we do not observe a strong gradient between the ejecta cloud and tail. Further, our polarimetric maps do not reveal any obvious features within the tail. This contrasts with the findings of \cite{opitom2023}, who observe that the tail generally appears redder in colour (associated with larger particles) compared to the surrounding bluer dust cloud (associated with smaller particles) in their so-called spectral-gradient maps. Again, the lack of a measurable polarisation gradient between the dust cloud and tail or within the tail could be due to the faintness, and hence low S/N, of the tail and/or the mixing of particles of various sizes. Further analysis is required to better investigate the polarisation of the dust tail. 

\begin{figure*}
\begin{center}
    \centering
    \includegraphics[width=14cm]{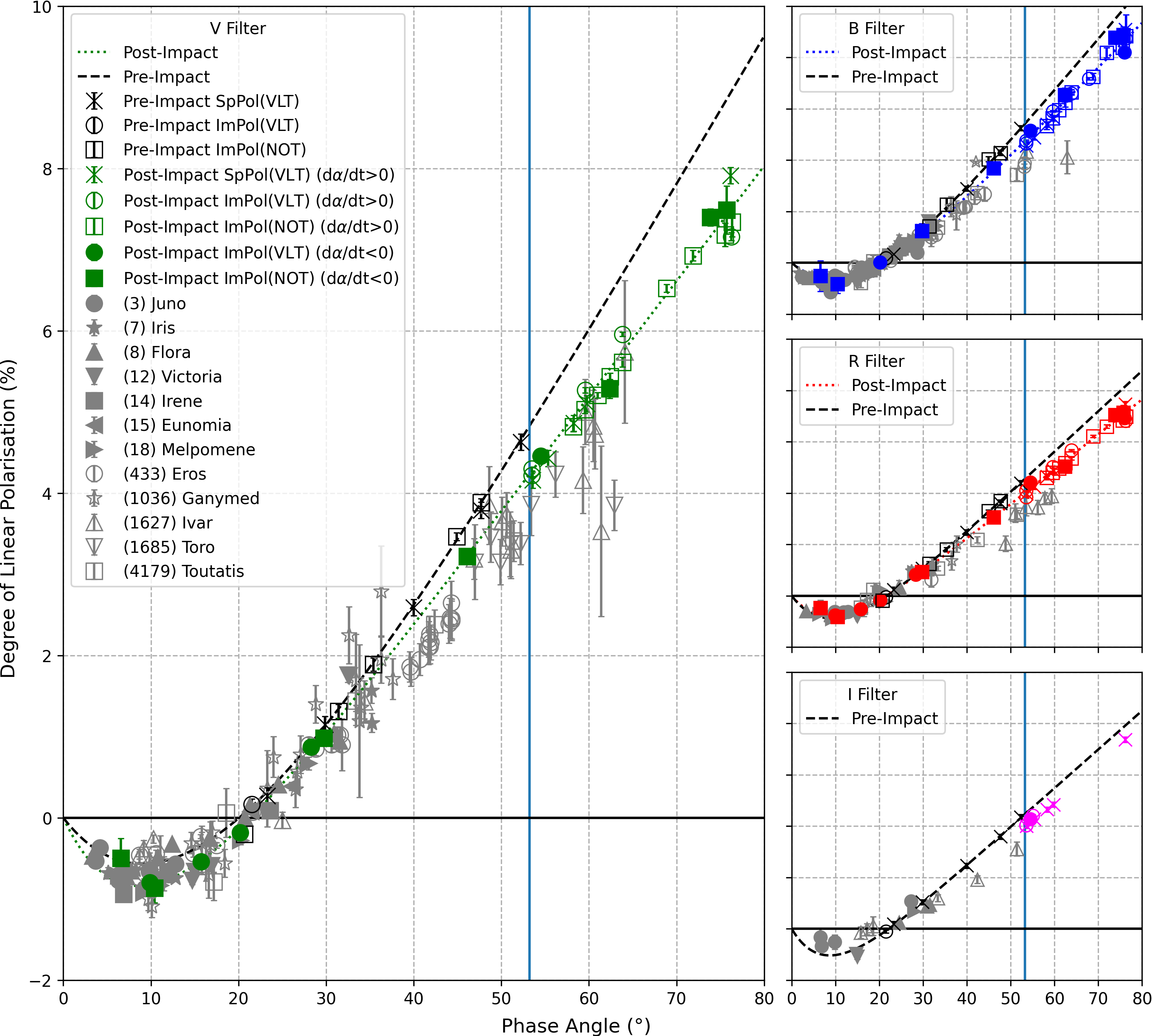}
    \caption{Polarimetric phase curve of Didymos compared to S-type asteroids. The main-belt asteroids are shown with filled, grey data points, while near-Earth asteroids are shown in open, grey points. Same as Figure \ref{fig:phasecurve}, the black points and curve represent pre-impact data, while the blue, green, red and pink points and fits represent the post-impact Didymos data (in the BVRI filters, respectively). The solid data points represent observations that were taken after the phase angle  peak was reached on Oct 20. In the plot for the V-filter, the majority of the Didymos-Dimorphos data points are values which have been converted from the R-filter using Eq. \ref{eq:r2v}.}
    \label{fig:astcomp}
\end{center}
\end{figure*}

\subsection{Comparison with Other Asteroids}
Based on previous observations, the physical characterization of Didymos-Dimorphos has posed a puzzling challenge. First, visible spectroscopic measurements by \cite{binzel2004} classified the system as an Xk-type. Later, spectroscopic measurements which extended into the near-IR range by \cite{deleon2006,deleon2010} classified it as an S-type. \cite{dunn2013} found that it is spectroscopically most consistent with L/LL-type ordinary chondrites, which are among the most common meteorites to fall to Earth. More recent observations by the James Webb Space Telescope (JWST) show that its spectrum is most consistent with ordinary chondrite meteorites and is similar to S-class asteroids \citep{rivkin2023}. Polarimetric studies play an important role in supporting and complementing taxonomic classifications based on reflectance spectra. The detailed morphology of the polarisation phase curve, including the values of $\pmin$, \amin (\pmax and \amax typically cannot be measured), $\ainv$, and the slope of the curve at \ainv (\textit{h}), are highly sensitive to the physical properties of each individual asteroid. Strong analogies are found in the polarimetric phase curves of asteroids belonging to the same taxonomic class and polarimetry has often proved useful in identifying objects of rare classes \citep[e.g.]{belskaya2005,cellino2006,gilhutton2008}.

For this study, we have compared the polarimetric phase curve of Didymos to that of other asteroids using data from literature, primarily that accumulated in the Asteroid Polarimetric Database (APD) \citep{apd2022} and the Calern Asteroid Polarisation Survey (CAPS) \citep{bendjoya2022}. Since most observations in these databases were performed in the V-band, while the majority of our observations were performed in the R-band, we converted our measurements from R to V with Eq. \ref{eq:r2v}, as discussed in Section \ref{sec:results_phasecurve}. Figure \ref{fig:astcomp} shows the polarimetric phase curve of Didymos compared to a number of S-type asteroids: main-belt asteroids (3) Juno, (7) Iris, (8) Flora, (12) Victoria, (14) Irene, (15) Eunomia, and (18) Melpomene; and  near-Earth asteroids (NEAs) (433) Eros, (1036) Ganymed, (1620) Geographos, (1627) Ivar, (1685) Toro, and (4179) Toutatis. In the V-filter panel of this plot, $P_Q(V_R)$ values were used in epochs were $P_Q(V)$ measurements were not obtained---i.e. the majority of the data points are those converted from the R-band to the V-band. The main-belt and NEAs have been distinguished with filled and open data points, respectively.

The polarimetric phase curve of Didymos-Dimorphos shows that it is generally in good agreement with other S-type asteroids. The additional data in the V band provided by Eq. \ref{eq:r2v} allow us to calculate a number of values in the polarimetric phase curve of the post impact data which are considered as indicative of the properties of the surface of asteroids. In particular, we measure $\pmin = (-0.84 \pm 0.10)\%$ at $\amin = (8.7 \pm 0.09)^{\circ}$, and $\ainv = (21.4 \pm 0.1)^{\circ}$ with a slope of $h = 0.11 \pm 0.01$, and hence albedo of $\rho_{\rm{v}} = 0.20 \pm 0.02$. These values are within the range of polarimetric phase curve parameters of S-type asteroids, which are typically characterised by $\pmin \sim -0.5$--$-1 \%$ around $\amin \sim 8^{\circ}$ and \ainv in the range $18-22^{\circ}$ \citep[e.g.]{gilhutton2008,lopez2019}. The albedo of Didymos-Dimorphos, both pre- and post-imapact systems, is in the brighter end of the typical albedo range of $0.1-0.28$ (mean $\sim 0.23$) for S-type asteroids \citep{demeo2013}. Further comparison of the pre- and post-impact system to other asteroids separately is challenging for a number of reasons. (i) The lack of pre-impact data at small phase angles means we cannot determine $\pmin$, an important quantity for distinguishing different taxonomic classes. Thus, we cannot assess the effect of the DART impact on the NPB. (ii) Most asteroids, other than NEAs, are only observable up to phase angles $\sim 30-35^{\circ}$. It is only beyond this point, however, that the pre- and post-impact fits diverge to a meaningful degree. For this reason, we are particularly interested in comparing Didymos-Dimorphos to other NEAs. (iii) The number of NEAs that have been observed in polarimetric mode at large phase angles, especially $>53^{\circ}$, is very small. 

Since Didymos-Dimorphos was intially classified as an Xk-type, we also compared the polarimetric curve to that of X-type asteroids. The polarimetric phase curves of X-types, however, show high dispersion across large phase angles ranges. This dispersion is likely due to the existence of many sub-classes of X-type asteroids, which include objects classified as E-, M-, and P-type in the \cite{tholen1989} taxonomy. Initially, these classes were described as spectrally degenerate due to the absence of mineral absorption features in their visible and near-infrared reflectance spectra. Instead, they were differentiated solely based on their albedo. As the shape of the polarisation phase curve is directly linked to the albedo, the polarisation of these objects follow distinct curves which depend on their taxonomic classification. Consequently, when plotted together, the resulting plot looks dispersed \citep{2012canada}. Although Didymos-Dimorphos shows a similar polarimetric phase dependence as a small selection of X-type asteroids (e.g. (92) Undina, (184) Dejopeja, (276) Adelheid, (1355) Magoeba and (2001) Einstein, all of which have been observed at phase angles $<30^{\circ}$), we cannot classify Didymos-Dimorphos as an X-type based solely on polarimetric data. On the other hand, we can rule out that Didymos does not belong to the C-type taxonomy, typically characterised by a much deeper NPB and steeper slope \citep{cellino2015b}, nor F-type, characterised by a smaller inversion angle around $\sim 15^{\circ}$ \citep[e.g. (3200) Phaethon]{devogele2018b}, or V-type, characterised by a shallower NPB \citep[(4) Vesta ]{lupishko1988}. We therefore conclude that, polarimetrically, Didymos-Dimorphos resembles S-type asteroids the most, despite the general scattering of data points at larger phase angles.

\section{Summary}
We monitored the Didymos-Dimorphos asteroid system in imaging polarimetric mode from around one month before, up to four months after the impact of the DART spacecraft. During this campaign, we have observed the system in three stages: (i) pre-impact, (ii) immediately post-impact, when the dust cloud is present, and (iii) long-term post-impact, when the dust cloud has dissipated and left behind an extensive dust tail. We performed our observations with FORS2 of the VLT in BVRI filters (mostly B and R) and ALFOSC of the NOT in B and R filters. A number of the VLT observations were performed hand in hand with spectropolarimetric observations also obtained with FORS2 \citep{bagnulo2023}. In total, we have obtained 30 observing series in the B filter, 4 in the V filter, 33 in the R filter and 5 in the I filter. Some of the observations could be used to check for a modulation of the polarisation with the rotation of Didymos, as well as with the orbit of Dimorphos. This study includes aperture polarimetric measurements, imaging and polarimetric maps, and hence, analysis of the polarimetric phase curve and spatial and temporal evolution of the ejecta dust cloud in polarised light. \citet{bagnulo2023} observed a drop of polarisation following the impact, but could not assess whether, after the ejecta cloud dissipated, the Dydimos-Dimorphos system came back to the polarimetric behaviour observed prior impact. Our new data allows us to establish the long-term, as well as the short-term, effects of the DART impact on Didymos-Dimorphos. 

We summarise our findings as follows:

\begin{enumerate}
    \item Immediate effects of DART: the most significant effect of the DART impact is the dramatic drop in polarisation observed immediately after impact. This drop suggests that the material ejected from the asteroid is different from that present on the original surface. As a possible interpretation, we suggest that the ejected particles are smaller and/or brighter than those on the pre-impact surface, in agreement with \cite{bagnulo2023}.
    \item Long-term effects of DART: the drop in polarisation persists as a long-term phenomenon, even after the majority of the dust cloud has dissipated. We suspect that this is due to residual material from the ejecta cloud still remaining in the system, either in orbit or deposited on the asteroid's surface. 
    \item Ejecta-cloud and tail: a large dust cloud is ejected immediately after the impact and persists for a number of weeks, eventually forming into an extensive tail. Several features are apparent in the imaging maps, but are not so obvious in the polarimetric maps. The dust cloud polarises light in an almost-uniform manner and a gradient is found between the polarisation in the dust cloud/tail and the surface of the asteroid.
    \item Variation in polarisation: some discrepancies among measurements taken at similar phase angles are observed. We find marginal evidence that these variations are linked to the rotation of Didymos, potentially implying compositional differences on the surface of Didymos. Further investigations are needed to confirm this possible finding. We have not found any correlation between these discrepancies and the eclipsing events of the two asteroids. 
    \item Polarimetric taxonomy: Didymos-Dimorphos resembles an S-type asteroid according to its polarimetric phase curve.
\end{enumerate}

In this paper, we provide quantitative measurements of the polarisation as well as qualitative explanations to its implications with regards to the properties of the dust (particle size and brightness). Future modelling studies will provide more quantitative estimates of these parameters. Future observations and analyses will also be essential to unravel the underlying causes of the observed discrepancies and to gain a more comprehensive understanding of the system behaviour. 

\section*{Acknowledgements}
This work is based on observations made with European Southern Observatory (ESO) telescopes at the La Silla Paranal Observatory under run ID 109.23GL and Director Discretionary Time with run ID 110.259G, as well as observations made with ALFOSC at the Nordic Optical Telescope (NOT) under proposal ID 66-020, which is provided by the Instituto de Astrofísica de Andalucía (IAA) under a joint agreement with the University of Copenhagen and NOT. NOT is owned in collaboration by the University of Turku and Aarhus University, and operated jointly by Aarhus University, the University of Turku and the University of Oslo, representing Denmark, Finland and Norway, the University of Iceland and Stockholm University at the Observatorio del Roque de los Muchachos, La Palma, Spain, of the Instituto de Astrof\'{\i}sica de Canarias. Research by Zuri Gray is funded by the UK Science and Technology Facilities Council (STFC). Geraint H. Jones is grateful to STFC for partial support through consolidated grant ST/W001004/1. Research by Karri Muinonen, Antti Penttil\"a, and Mikael Granvik is supported by the Academy of Finland (Research Council of Finland) grants No. 345115 and No. 336546. L. Kolokolova acknowledges support from NASA DART PS Program, grant no. 80NSSC21K1131. \\

We are grateful to the referees for their thorough examination of the manuscript and their insightful comments and suggestions. 

\bibliography{ZGreferences}{}
\bibliographystyle{aasjournal}

\end{document}